# Vaporization of the Earth: Application to Exoplanet Atmospheres

Laura Schaefer[1,2]

Katharina Lodders[1]

Bruce Fegley, Jr.[1]

[1]Planetary Chemistry Laboratory, Dept. of Earth & Planetary Sciences and McDonnell Center for the Space Sciences, Washington University, St. Louis, MO 63130 USA

laura_s@wustl.edu, bfegley@wustl.edu, lodders@wustl.edu

[2]Present address: Harvard/Smithsonian Center for Astrophysics, 60 Garden St, Cambridge, MA 02138 USA

lschaefer@cfa.harvard.edu

Running Title: Vaporization of the Earth

Tables: 4

Figures: 11

**ABSTRACT**

Currently, there are about 3 dozen known super-Earth ($M < 10\ M_{\oplus}$), of which 8 are transiting planets suitable for atmospheric follow-up observations. Some of the planets are exposed to extreme temperatures as they orbit close to their host stars, e.g., CoRot-7b, and all of these planets have equilibrium temperatures significantly hotter than the Earth. Such planets can develop atmospheres through (partial) vaporization of their crustal and/or mantle silicates. We investigated the chemical equilibrium composition of such heated systems from 500 – 4000 K and total pressures from $10^{-6}$ to $10^{+2}$ bars. The major gases are $H_2O$ and $CO_2$ over broad temperature and pressure ranges, and Na, K, $O_2$, SiO, and O at high temperatures and low pressures. We discuss the differences in atmospheric composition arising from vaporization of $SiO_2$-rich (i.e., felsic) silicates (like Earth's continental crust) and MgO-, FeO-rich (i.e., mafic) silicates like the bulk silicate Earth. The computational results will be useful in planning spectroscopic studies of the atmospheres of Earth-like exoplanets.

**KEYWORDS: astrochemistry; atmospheric effects; planets and satellites: general**

## 1. INTRODUCTION

Observations of the atmospheric chemistry of some super-Earth exoplanets ($M<10\ M_{\oplus}$) are now possible. Transmission spectra of GJ 1214 b, first discovered by Charbonneau et al. (2009), were measured by Bean et al (2010, 2011), and subsequently Désert et al. (2011), Croll et al. (2011), Crossfield et al. (2011), and Berta et al. (2012). Table 1 lists the currently known exoplanets with measured radii and measured masses under $10M_{\oplus}$, ordered by the calculated equilibrium temperatures from stellar irradiation. The Kepler planets and 55 Cnc e orbit G stars similar to the Sun, whereas CoRoT-7 b orbits a slightly cooler K star and GJ 1214 b orbits a much cooler and smaller M star. These planets range in density from 0.71 to 10.4 g cm$^{-3}$, with equilibrium surface temperatures greater than 500 K. In comparison, the Earth's bulk density is 5.52 g cm$^{-3}$, with an equilibrium temperature of ~255 K. The densities listed in Table 1 suggest that these super-Earths have a wide range of bulk compositions, from H/He-rich like Saturn (density of 0.687 g cm$^{-3}$), to possible "water" worlds, or even super-Mercuries. However, determining planet compositions based solely on bulk densities is not possible; for a given bulk density, there can be significant degeneracies in the bulk composition based on mixtures of gas ($H_2$), water/ice, rock, and metal. For instance, possible bulk compositions for GJ 1214 b include a planet with a substantial H/He envelope, or a water or ice world with a steam atmosphere (Rogers & Seager 2010; Nettlemann et al. 2011). Atmospheric models of GJ 1214 b suggest that spectroscopic observations should be able to distinguish between low mean molecular weight H/He atmospheres with large scale heights and those with a heavier composition such as a steam or $CO_2$ dominated atmosphere (Miller-Ricci & Fortney 2010).

The example of GJ 1214 b shows atmospheric observations can help to establish the bulk composition of a planet. However, models of atmospheric chemistry are essential to decipher the observed spectra by predicting likely compositions, and pinpointing tracer molecules that may be easily identified in various spectral regions. In this paper, we focus on modeling the vaporization of an Earth-like planet. All currently known transiting super-Earths have temperatures significantly higher than the Earth's. Thus, it is important to develop models of the possible atmospheric compositions of Earth-like planets at high temperatures. Past models have generally focused on either broad composition classes for outgassed planetary atmospheres (water, water+$CO_2$, $H_2$, etc.) (e.g., Elkins-Tanton & Seager 2008; Miller-Ricci & Fortney 2010) or on models for planets in the habitable zone (HZ) based on the present day terrestrial atmosphere (e.g., Segura et al. 2003, 2005). A more detailed compositional model for outgassed atmospheres will help predict possible spectroscopically important, albeit less abundant gases, while addressing the question of what the Earth's atmosphere would look like sans life. The chemical equilibrium calculations here are not meant to describe any particular known exoplanet, but to map the atmospheric chemistry of an Earth-like planet across a wide temperature-pressure space. This effort will help identify major and minor gas species that may be of interest for characterizing super-Earth atmospheres. Some known exoplanets may fall within the temperature-pressure space studied here, and where this occurs, we can state what atmospheric composition would be expected for the planet at these conditions, were it to be of an Earth-like composition. As actual observations of these atmospheres become possible, comparison of predicted Earth-like compositions to the actual observed atmospheres may be useful to determine the oxidation state of these planets' interiors.

This paper is organized as follows. In §2, we describe our thermochemical equilibrium calculations. In §3, we describe the compositions of the atmospheres formed by outgassing of Earth-like planets with a crust like Earth's continental crust (composed of $SiO_2$-rich silicates) and those which have not formed a felsic crust and have a silicate lithosphere like Earth (composed of MgO-, FeO-rich silicates). We also discuss the variation of gas chemistry for a range of total atmospheric pressures and volatile element abundances. In §4, we discuss some applications of our calculations to exoplanets. We also make recommendations for species that could be detected in the atmospheres of evaporating super-Earths, and briefly discuss the effects of photochemistry on the atmospheric composition. Preliminary results of this work were given by Schaefer et al. (2010, 2011). Schaefer and Fegley (2007, 2010) previously modeled outgassing of chondritic composition material and we refer the reader to these papers for a discussion of the atmospheric chemistry expected for chondritic planets.

## 2. COMPUTATIONAL METHODS

We used thermochemical equilibrium calculations to model atmospheric compositions formed by rocky planets similar to the Earth that are subjected to high temperatures. In order to do this, we must chose a reasonable representative composition of the silicate portion that would be subject to heating and responsible for creating an atmosphere through outgassing. We use two different rocky compositions: (1) the continental crust (Wedepohl 1995), and (2) the bulk silicate Earth (BSE) (Kargel & Lewis 1993), which are taken here to represent a good approximation to rocky planetary compositions. These compositions allow us to examine differences in atmospheres outgassed by $SiO_2$-rich (felsic) silicates (like Earth's continental crust) and MgO-, FeO-rich (mafic) silicates (like the bulk silicate Earth). For reference, the bulk silicate Earth (BSE) is the composition of the Earth's silicate portion that evolved into the atmosphere, oceans, crust, and mantle. Because the mantle comprises 99.4% by mass of the BSE, the BSE composition is similar to that of Earth's mantle.

We can expect that any rocky planet is differentiated into a silicate portion and a more dense metallic core. Of course, the chemical composition of any rocky planet will depend on the conditions of the planetary accretion disk in which it formed and the volatile contents of the planetesimals from which it accreted. However, elements that are volatile in a solar-composition gas show strong depletions in all rocky planets within the solar system, indicating that the planetesimals accreting to terrestrial planets never contained a full solar complement of volatile elements. The same is true for differentiated meteorites which are remnants of small worlds that differentiated into silicate mantles and metallic cores. This depletion of volatile elements is a well-known characteristic of differentiated objects, as documented in the cosmochemical literature (e.g., numerous references in Lodders & Fegley 1998, 2010). Furthermore, differentiation occurred very rapidly and was likely concurrent with planetesimal and planet formation (i.e., within one Myr of the formation of the first solids in the solar nebula; e.g. Burkhardt et al. 2008). This also affects the silicate composition. Thus, a H- and noble gas-free but otherwise full solar composition for the rock forming elements would not make a good proxy for the evaporating silicate portion of a planet, as it would contain too much C, N, O, and moderately volatile elements with respect to the refractory elements. In addition, one would need to adjust the silicate composition for the depletion of Fe, Ni and other elements that are sequestered into a metallic core. This, however, largely depends on the oxidation state (including the C/O ratio) and would require a further set of modeling assumptions. One could make arbitrary adjustments to the compositions and assume various degrees of depletions of volatile rock-forming elements, as well as various estimates for element sequestration into a dense core, which would not be included in the atmosphere formation process, but this is far too much compositional space to explore at the start. Hence we decided to start our exploration of atmospheric compositions with silicate composition that are well studied. However, we have also explored some aspects of variable H, C, and O contents in §3.2 since magmatic contents of these elements may vary, and they determine the major composition of an outgassed atmosphere at lower temperatures.

One could ask if nebular H and He rich gas accretion to a hot rocky planet would alter the composition of the atmosphere and thus the conclusions of this paper, in which only evaporative formation of an atmosphere is assumed. Accretion of H and He rich gas can only happen by (a) accretion of solid planetesimals containing H and He; or (b) by capture of nebular gas during the typically short (e.g., less than 10 Ma) lifetime of the nebula disk. In case (a) any (unlikely) gain of H and He during planetesimal accretion would be offset by loss of H and He by atmospheric escape from Earth-sized and smaller planets. Venus is an obvious and well-studied example that H-loss is unavoidable, and Venus' surface temperature of 470 K is also much lower than the magmatic temperatures employed for the hot rocky planets here. More massive planets may be able to retain some H-rich gas, particularly if they are cooler planets and not subjected to intense heating, such as GJ 1214 b. In this paper, we focus more on the hot planets with compositions similar to that of the Earth (i.e., poor in H and He); however, we discuss differences in atmospheric chemistry caused by variations in the volatile elements in Section 3.2. For case (b), the planet must be massive enough to capture H and He and to retain it against atmospheric escape. Such planets fall into the Neptune-like category and are not discussed here. We comment on some of these aspects in more detail below.

Our calculations were done with a Gibbs-energy minimization code of the sort described by Van Zeggern & Story (1970). Table 2 gives the elemental composition of the two models by mass percent. We considered ~750 gas species and ~60 solid and liquid compounds of the 18 elements included in our calculations. We considered most abundant elements except Ni (~0.18%), which alloys with Fe and has very similar chemistry. Our discussion of results focuses on the continental (felsic) crust with important differences for the BSE being noted as necessary. We do this because formation of continental (felsic) crust requires water and the most interesting Earth-like exoplanets are those with abundant water. We discussed the atmospheric chemistry of rocky, water-depleted (i.e., Venus-like) exoplanets in another paper (Schaefer and Fegley 2011).

For our calculations, the temperatures range from 500 K to 4000 K. All currently known transiting super-Earths (see Table 1) have equilibrium temperatures from stellar irradiation that fall within this range. We considered total pressures from $10^{-6}$ bars to $10^{+2}$ bars. Low pressures are applicable to volatile-poor planets or planets with transient atmospheres, which are being actively removed by some mechanism. High pressures are applicable to planets with abundant volatiles (e.g., like Earth and Venus). We discuss the variations in the gas composition with pressure for both the continental crust and the BSE. We further consider differences in the atmospheric composition for variations in the abundances of H, C, and O, as discussed in §3.2.

## 3. RESULTS

### *3.1 Atmospheric Composition at 100 bars*

#### *3.1.1 Continental Crust*

Figure 1 shows the composition of the gas phase at 100 bars as a function of temperature for the continental crust. We only show gases with abundances greater than 0.1% by volume (mole fraction $X_i > 10^{-3}$) for clarity. The graphs show the major gases of (a) the volatile elements H, C, N, and S, and (b) lithophile (or rock-forming) elements, such as Na, K, Fe, Si, Mg, Ca, Al, and Ti. The atmosphere formed by outgassing of the continental crust is dominated over a very wide temperature range by ~55% steam and ~42% $CO_2$. Carbon is 100% in the gas phase down to temperatures of ~700 K, where it forms carbonates such as $CaMg(CO_3)_2$ (dolomite). Carbon forms numerous minor gases, but the bulk of the carbon is in $CO_2$ from ~700 – 3000 K. At higher temperatures, much of the carbon goes into CO gas, which has a peak abundance of ~7% at 3750 K.

Between 3450 – 3900 K, the dominant atmospheric species is $O_2$ gas. Diatomic oxygen forms by thermal dissociation of $H_2O$, $CO_2$, and $SO_2$, and is also released by vaporizing silicates. It has abundances > 0.1% at temperatures above ~1600 K. As temperature increases, the mole fraction of $O_2$ plateaus and drops off at the highest temperatures. However, the percentage of oxygen residing in $O_2$ continues to increase as $H_2O$ and $CO_2$ dissociate. The $O_2$ mole fraction curve in Figure 1 does not illustrate this increase clearly due to the increasing abundance of material in the gas phase with increasing temperature. As silicates vaporize, significant amounts of rock-forming elements go into the gas phase, thereby decreasing the relative proportions of other gases such as $O_2$. We will discuss this phenomenon in greater detail when we discuss Fig. 3. Monatomic O is formed by thermal dissociation of the major O-bearing gases and by vaporization of silicates. It is abundant above ~2500 K, with a peak abundance of 6.6% at 4000 K. Large amounts of O and $O_2$ are formed by silicate vaporization at high temperatures because silicates vaporize incongruently. This means that silicates do not vaporize to gas molecules of the same composition but instead dissociate and form atoms and simpler molecules during vaporization. Schaefer and Fegley (2004, 2009) previously noted the importance of O and $O_2$ in high temperature silicate vapor. The hydroxyl radical OH forms by thermal dissociation of water vapor. Hydroxyl is abundant at temperatures above ~2000 K and has a peak abundance of 6.9% at 3750 K. At higher temperatures, OH thermally dissociates to form monatomic H and O. Monatomic H has a peak abundance of 1.1% at 3900 K.

At low temperatures (< 600 K), the atmosphere is dominantly $N_2$. Nitrogen forms no solid compounds in this system, and is 100% in the gas phase at all temperatures. The major N-bearing species is $N_2$ up to ~3200 K. The survival of $N_2$ to ~ 3200 K is due to its large bond energy. At higher temperatures, dissociation of $N_2$ and reaction with oxygen forms NO gas. At 4000 K, NO gas is the dominant N-bearing compound, with an atmospheric abundance of ~1%. Sulfur is also a major component of the atmosphere. It forms several solid sulfides and sulfates at low temperatures but enters the gas in significant amounts at ~1000 K as $SO_2$, which is the dominant sulfur-bearing compound from 1000 – 4000 K, with a peak abundance of ~5% at 2250 K. At temperatures above 3000 K, $SO_2$ begins to dissociate to form significant amounts of SO (~10% of total sulfur at 4000 K).

The hydrogen halides HCl and HF are the dominant Cl- and F-bearing gases at lower temperatures while KCl and KF are the major Cl- and F-bearing gases at higher temperatures. The HCl/KCl ratio reaches unity at lower temperature (~ 2095 K) than the HF/KF ratio (~ 3530 K) because the dissociation energy of HCl is lower than that of HF. Bromine and iodine are trace elements in the continental crust and were not considered in our calculations.

At high temperatures, vaporization of silicates alters the atmospheric chemistry, as shown in Fig. 1b. Silicon is found primarily in silicates. At 4000 K, only ~10% of Si is in the gas phase, dominantly as SiO, followed by $SiO_2$. At 4000 K, SiO gas is the major gas in the whole atmosphere, with a maximum abundance of ~29%. Schaefer and Fegley (2004, 2009) previously noted the importance of SiO in high temperature silicate vapor. Other lithophile (i.e., rock-forming) elements are also in the atmosphere at temperatures > 2000 K. The most abundant of these are the alkalis Na and K, which have maximum abundances of 4% and 3.5%, respectively. Alkali chlorides (e.g. KCl, NaCl, etc.) are the major Na- and K-bearing gases at temperatures below ~2000 K. Above 2000 K, the alkali hydroxide gases become more abundant than the alkali chlorides. Iron is also abundant in the gas phase. Iron hydroxide $Fe(OH)_2$ is the major Fe-bearing gas with a peak abundance of ~0.2% ($T < 3500$ K). At higher temperatures, dissociation of $Fe(OH)_2$ produces monatomic Fe and FeO gas, which have abundances of 0.4% and 0.6%, respectively, at 4000 K. MgO gas is the major Mg-bearing gas and has an abundance of ~0.3% at 4000 K. At 4000 K, roughly 61% of all titanium is in the condensed phase, with the remaining 39% in the gas phase. Titanium dioxide $TiO_2$ is the major Ti-bearing gas at all temperatures studied with an abundance of 1.1% at 4000 K. Phosphorus, manganese, and chromium gases are not shown to simplify the graph because these three elements have complex chemistry. Phosphorus gases start to become important at ~ 2500 K and the major P-bearing gas shifts from $P_2O_5$ to $PO_2$ to PO with increasing temperature. The bulk of Mn and Cr are vaporized at higher temperatures and with increasing temperature their major gases are $MnF_2$, MnO, and Mn; and $CrO_2F$, $CrO_2$, and CrO. Calcium and aluminum are not abundant enough in the gas phase to appear in Figure 1. Their major gases are $Ca(OH)_2$ and AlO, which have maximum abundances of 2 and 800 ppmv (parts per million by volume), respectively.

*3.1.2 Bulk Silicate Earth*

Figure 2 shows the results of our computations for the BSE. As with Fig. 1, the figure is split into (a) volatile and (b) rock-forming elements. As for the continental crust, the atmosphere of the vaporized BSE is dominated over a wide temperature range by $H_2O$ and $CO_2$, although the relative abundance of $CO_2$ (~16%) is significantly lower. At low temperatures, the atmosphere of the BSE is significantly more reduced than the continental crust, and is dominated by methane and ammonia. The reason for this is that Earth's continental crust contains ferric iron ($Fe^{3+}$) minerals and is more oxidized than the bulk silicate Earth, which contains mainly ferrous iron ($Fe^{2+}$) minerals. Methane is more abundant than $CO_2$ below ~750 K, and is the most abundant gas in the atmosphere, whereas ammonia $NH_3$ is more abundant than $N_2$ below 650 K. The abundances of minor gases such as $SO_2$ (~20% at 1800 K) and $H_2$ (~4% at 700 K) are significantly larger in the BSE atmosphere. At higher temperatures, $O_2$ becomes the dominant gas at ~2900 K, compared to ~3400 K for the continental crust.

Rock-forming elements in general are more abundant at high temperatures in the BSE atmosphere than in the continental crust, but the abundances of SiO (~2.2%) and $SiO_2$ (~0.6%) are much lower than for the continental crust. The most abundant lithophile gas at high temperatures is monatomic Na, with a maximum abundance of ~23% at 3850 K. The alkali hydroxides KOH and NaOH are also fairly abundant, with peak abundances of ~7% at 2900 K and 3450 K, respectively. Magnesium is more abundant in the BSE atmosphere than above the continental crust. At temperatures less than ~3500 K, Mg is mostly found in $Mg(OH)_2$, which has a peak abundance of 0.3% at 3400 K. At higher temperatures, dissociation of $Mg(OH)_2$ produces MgO gas, which has a maximum abundance of 4% at 4000 K. Titanium and iron behave similarly in the BSE and the continental crust, with abundances of ~1% $TiO_2$ and FeO at 4000 K. Aluminum and calcium are again not abundant enough to appear on the graph, with maximum abundances of 20 ppmv CaO and 600 ppmv AlO.

*3.2 Variable Abundances: H, C, O*

To explore the effect of variations in the element abundances, we did a series of calculations at 100 bars for both the continental crust and the BSE with variable abundances of H, C, and O. Calculations were done from 0.1 – 10× the nominal abundances for C and H, for both the continental crust and the BSE. As discussed below, there is no evidence that the actual C and H abundances are so different from our adopted nominal values. However we have done calculations over these wide ranges to illustrate trends. The calculations for variable oxygen abundances were done using a smaller range of abundances as described later. Each element was varied individually, with all other elements remaining fixed at their nominal abundances.

Earth and other rocky bodies (e.g., the Moon, Mars, 4 Vesta) are O-rich and C-poor and have C/O atomic ratios that are orders of magnitude smaller than the solar composition value of ~ 0.46 (Lodders and Fegley 2010). The C/O atomic ratios in Earth's crust and the bulk silicate earth are 0.0056 and 0.0002, respectively (Table 2). Geochemical analyses of lunar rocks, the Martian SNC meteorites, and the EHD meteorites (from 4 Vesta) show similarly small C/O ratios on these bodies.

Silicate exoplanets are also expected to have C/O ratios significantly less than the solar composition value because the major condensates formed from solar composition material are O-rich and C-poor due to the instability of carbides, carbonates, and graphite in solar composition material (Lewis et al. 1979), and the stability of O-rich silicate condensates such as enstatite ($MgSiO_3$) and forsterite ($Mg_2SiO_4$) in solar composition material (e.g., Larimer 1967, Barshay & Lewis 1976, Lodders 2003, 2009).

Condensation calculations as a function of variable C/O ratio also show that the only "rocky" bodies that can possibly have C/O ratios approaching unity would be formed at C/O ~ 1 in otherwise solar composition material (e.g., see Larimer 1975, Larimer & Bartholomay 1979, Lodders & Fegley 1997). Thus the situation for rocky planets is completely different than that for gas giant planets where C/O ratios higher than the solar composition value can be produced by formation from solar composition material, e.g., as discussed by Lodders (2004) for Jupiter.

*3.2.1 Hydrogen*

We use 0.045 wt. % H as the hydrogen abundance in Earth's continental crust (Table 1) from Wedepohl (1995). Li (1972) considered geochemical mass balance during rock weathering and derives 2.9 wt. % $H_2O$ in sedimentary rocks ($2.4 \times 10^{21}$ kg total mass). His value corresponds to 0.051 wt. % H in Earth's continental crust, and is close to our adopted value. Hunt (1972) computed $9 \times 10^{17}$ kg H bound with reduced carbon in sedimentary rocks, corresponding to 0.0059 wt. % H in the continental crust. This is 7.6 times smaller than our adopted value, but does not include water in sediments.

Our calculations with variable H abundance for the continental crust and BSE are shown in Figure 3a and b, respectively. The middle panels show the nominal abundances (see Fig. 1 & 2). Rocky element gases are not shown, but described where appropriate. Reducing the hydrogen abundance of the continental crust causes the $H_2O$ gas abundance to drop, becoming the third most abundant gas at low temperatures. The first and second most abundant gases are then $CO_2$ (500-2600 K), and $O_2$ (2600 – 4000 K). Abundances of other H-bearing gases (OH, H, HF, HCl, KOH, NaOH) drop slightly, but non-H-bearing gases are relatively unaffected. For the BSE, lowering the H abundance similarly lowers the abundances of the H-bearing gases, so that the most abundant gases are $CH_4$ (500-700 K), $CO_2$ (700-1800 K), $SO_2$ (1800-2900 K), and $O_2$ (2900-4000 K). Increasing the hydrogen abundance increases the abundances of all H-bearing gases, such as $H_2$, $H_2O$, $CH_4$, OH, H, and KOH. For the continental crust, $H_2$ is the most abundant gas at all temperatures (~45 vol%), followed closely by $H_2O$ (~33 vol%), whereas the reverse is true for the BSE ($H_2$~28%, $H_2O$~52%). These gases make up nearly 80% of the gas. For both the continental crust and the BSE, the third most abundant gas is $CH_4$ (500-1200 K), or CO (>1200 K). Also, $CO_2$ is less abundant than CO for both the continental crust and the BSE, which is a major change from the nominal gas compositions. For the continental crust, the abundance of SiO increases and is greater than KOH above ~2800 K (compared to ~3700 K for the nominal composition). For the BSE, Na and NaOH have about the same abundances, but neither becomes the most abundant gas at high temperatures.

*3.2.2 Carbon*

We adopt 0.199 wt. % carbon in Earth's continental crust (Table 1) from Wedepohl (1995). Most of this (~ 80 %) is in sedimentary rocks as carbonates and reduced carbon (e.g., coal, oil, natural gas, kerogen) in a 4:1 mass ratio, and the other 20% is in igneous and metamorphic crustal rocks. The relative amounts of carbonates and reduced carbon in the crust are constrained by their isotopic compositions of $\delta^{13}C = 0$ ‰ (carbonates) and $\delta^{13}C$ = -25 ‰ (reduced carbon) as discussed by Wedepohl (1995). Older values for the carbon content of Earth's continental crust are up to 3 times larger than Wedepohl's value and are $5.5 \times 10^{19}$ kg carbon (0.36 wt. %) from Junge et al. (1975) and $9 \times 10^{19}$ kg carbon (0.59 wt. %) from Hunt (1972).

Calculations with variable carbon abundances are shown in Figure 4. For a reduction in the carbon abundance, the abundances of the C-bearing gases drop, but relative abundances of non-C-bearing gases remain relatively unaffected for both the continental crust and the BSE. For the continental crust, the $CO_2$ abundance drops from ~30% to 3%, whereas for the BSE, $CO_2$ drops from ~9% to 1%. Lithophile gas chemistry is relatively unchanged. Increasing the carbon abundance, on the other hand, alters gas chemistry more significantly, particularly for the continental crust. For the continental crust, the two most abundant gases are either $CO_2$ and $H_2O$ (600 – 1100 K), or CO and $H_2$ (1100 – 4000 K). The abundances of $SO_2$ and $O_2$ are less than 0.1% until temperatures of ~3500 K, whereas the abundance of $CH_4$ is above 0.1% from 500 – 1600 K. In the lithophile gases, the abundances of the compound alkali gases (e.g. KOH, NaOH, KCl, NaCl), decrease, whereas the abundances of SiO, Na, and K are significantly larger than for the nominal case. SiO gas is the most abundant gas above ~3700 K. For the BSE, raising the C abundance by 10x increases the $CO_2$ abundance from 9% to ~44%. The major gases are $CH_4$ (500 – 700 K), and $CO_2$ (700 – 3700 K). The CO abundance also increases by about 10×, while the $H_2O$ and $O_2$ abundances decrease ($O_2$ peak abundance of ~10 % compared to 40% for the nominal case). The major gas above ~3700 K is monatomic Na, however, the lithophile gas chemistry is relatively unchanged from the nominal composition.

*3.2.3 Oxygen*

The abundance of oxygen in the BSE is difficult to determine, and is usually calculated assuming standard oxide compositions for the major elements (e.g., silicon is found as $SiO_2$, Al as $Al_2O_3$ etc.). The uncertainties in the abundances of the major elements and of the $Fe^{2+}/Fe^{3+}$ ratio in the BSE lead to a range of possible oxygen abundances from as low as 38.7% to as high as 48.2% by mass, in comparison to the average value of 44%. This is a difference of roughly ±10% of the nominal value used in calculating Fig. 2. The abundance of oxygen in the continental crust is better constrained because the major element abundances can be more tightly constrained by sampling and the $Fe^{2+}/Fe^{3+}$ ratio is much better known. Rudnick & Gao (2003) compiled and reviewed numerous analyses of the composition of the continental crust, and found a range of oxygen abundances of 45.5 – 47.1 wt%, with an average of 46.5 wt%. Note that our chosen continental crust composition (Wedepohl 1995) has a higher O abundance of 47.2 wt%, however Rudnick & Gao (2003) did not include volatile elements such as H and C, which commonly occur in oxidized form in the Earth's crust. Regardless, the range of possible oxygen abundances in the continental crust is far smaller than in the BSE. However, in Figure 5, we chose to show comparable ranges in oxygen abundance for the continental crust (47.2 ± 4 wt %) and the BSE (44.42 ± 4 wt%), for illustrative purposes.

Figure 5 shows the variable O-abundance calculations. For the continental crust, lowering the oxygen abundance to 43 wt% changes the major gas chemistry significantly. The most abundant gas is $H_2$ from 500 – 2800 K, above which the major gas is SiO. The second most abundant gas is $CH_4$ from 500 – 1800 K, and CO from 2000 – 2800 K. Increasing the O abundance to 51.2 wt% also strongly affects the gas composition. Dioxygen $O_2$ becomes the most abundant gas at all temperatures, and the abundances of $H_2O$ and $CO_2$ drop to ~14% and 10% respectively. For the BSE, assuming the lowest possible O abundance (not shown), we found an atmosphere composed of primarily $H_2$ from 500 – 2000 K, Na from 2000 – 3000 K, and SiO at the highest temperatures. Oxygen abundances from 39-42% maintained this pattern, with $H_2$ being the dominant gas from 500-2000 K, and Na at temperatures above 2000 K. In these calculations, the abundance of $H_2O$, $O_2$, and $SO_2$ never rose above $10^{-3}$. For smaller oxygen depletions (O>42.5%), $H_2$ remained the major gas over a wide temperature and pressure range, but $H_2O$, CO, and $CH_4$ were present at the percent level. For bulk O abundances of greater than 46%, the atmosphere was composed of ~100% $O_2$ up to temperatures of 3500 K. For an oxygen abundance of 45%, only 0.6% above the nominal value, the gas was more similar in composition to Fig. 2, with the major gases being $N_2$ (500-700 K), $CO_2$ (700-1000 K), $H_2O$ (1000 – 2000 K), and $O_2$ (2000 K).

### *3.3 Effects of Total Pressure on Gas Composition*

Table 3 summarizes the results for the major gas species at four temperature and pressure points for both nominal composition models. At both low temperature points and the high temperature and pressure point, water vapor is the dominant gas for both the continental crust and the BSE. The BSE generally produces a higher fraction of water vapor, and also significant amounts of $SO_2$. In contrast, the continental crust has only slightly more water vapor than $CO_2$, and contains relatively large amounts of hydrogen halide gases (HCl and HF). The high T and low P point (2000 K, $10^{-6}$ bars) has a much more unique gas composition, consisting mostly of SiO, O, and $O_2$, and also Mg gas for the BSE composition.

Figure 6 shows the gas to solid molar ratio as a function of temperature from $10^{-6}$ to 100 bars for (a) the continental crust and (b) the BSE. At low temperatures, the BSE has roughly an order of magnitude less material in the gas phase than the continental crust. This is because the BSE composition (mainly $Mg_2SiO_4$) is more refractory than the continental crust composition (mainly $SiO_2$ and feldspar). The nearly vertical increase in the gas/solid molar ratio occurs at the temperatures at which the condensed phase vaporizes. This temperature increases with total pressure, so that at the lowest pressure ($10^{-6}$ bars), nearly all material is vaporized at temperatures greater than ~1800 K, whereas at the highest pressures (100 bars), nearly all of the mass of the system remains in the condensed phase even up to the highest temperatures.

Figure 7 shows a comparison of the molar abundances of four major gases for both the continental crust and the BSE as a function of temperature for pressures ranging from $10^{-6}$ to $10^2$ bars. The abundance curves for the continental crust and BSE compositions are qualitatively similar for each volatile. However, there are differences in several details. The $H_2O$ abundances for the BSE have higher maximum values (but decrease more with increasing temperatures) than the $H_2O$ curves for the continental crust. The same behavior occurs for $SO_2$. In contrast, the $CO_2$ and SiO curves for the continental crust have higher maxima than those for the BSE. As pressure increases, the peak abundances and the decline of abundances for each gas shifts to higher temperatures. The temperature range over which the gas maintains its peak abundance also increases at higher pressures. For instance, $CO_2$ maintains a relatively constant abundance of about 30% (molar) up to temperatures of 1150, 1300, 1550, 1950, and 2300 K for pressures of $10^{-6}$, $10^{-4}$, $10^{-2}$, $10^0$, and $10^2$, respectively. The primary difference in gas abundance due to composition is for $SO_2$ gas, which is far more abundant in the BSE.

$SO_2$ gas should be spectroscopically observable in super-Earth atmospheres at concentrations greater than a few ppmv (Kaltenegger & Sasselov 2010), and may indicate a planet dominated by a global sulfur-cycle (like Venus, see Fegley 2004).

Figure 8 shows the most abundant compounds for the volatile elements H, C, N, and S as a function of pressure and temperature. The regions are the areas where the individual compounds dominate and lines indicate equal molar abundances of two neighboring species. The solid lines are for the continental crust and dashed lines are for the BSE. For example, water vapor is the major H-bearing compound in the field labeled $H_2O$, the hydroxyl radical is the major H-bearing compound in the field labeled OH, and monatomic hydrogen gas is the major H-bearing compound in the field labeled H. Molecular $H_2$ is never the major H-bearing gas for either composition. Models by Elkins-Tanton & Seager (2008) predict significant $H_2$ outgassing. However, these authors did not consider chemical equilibrium, but instead used models assuming either complete iron oxidation or no oxidation by $H_2O$. In contrast, our model, which deals with near-surface layers, does not constrain the total mass of the outgassed material and allows H and O to equilibrate with the BSE (or CC) and in the atmosphere.

Carbon dioxide is the major C-bearing species at low temperatures and high pressures. It dissociates to CO at lower pressures and higher temperatures. The BSE has a region of $CH_4$ dominance at low temperatures that does not occur for the continental crust. Nitrogen chemistry is dominated at lower temperatures by $N_2$, and by NO and N at higher temperatures. The BSE has a small region at low temperatures where $NH_3$ is the dominant form of N. Sulfur dioxide is the major S-bearing compound over a large P,T range. It dissociates at higher temperatures and lower pressures into SO, which itself dissociates at higher temperatures to monatomic S. At low temperatures, sulfur chemistry is different for the BSE and continental crust. Hydrogen sulfide gas is the major S-bearing compound for the BSE while sulfides and sulfates are the major S-bearing compounds for the continental crust.

## 4. ATMOSPHERIC CHEMISTRY

### *4.1 Discussion of Results*

In general the atmospheric composition predicted by our calculations agrees with what one expects from heating up the Earth's crust and vaporizing the volatiles it contains. Most crustal carbon is in the form of carbonates (~ 80% by mass) with a smaller amount (~ 20 mass %) present as reduced carbon (as coal, peat, natural gas, and oil). Heating up the carbonates will release $CO_2$ and oxidation of the reduced carbon during heating yields $CO_2$ and $H_2O$. Water vapor also results from vaporizing pore water and hydrated minerals in the crust. Very little CO or $CH_4$ is expected from heating the Earth's oxidized crust because there is no metallic iron to reduce $CO_2$ via reactions such as

$$4CO_2\ (g) + 3Fe\ (metal) = Fe_3O_4\ (magnetite) + 4CO\ (g) \quad (1)$$

In fact oxidation of CO and $CH_4$ by hematite ($Fe_2O_3$) will occur via reactions such as

$$CO + 3Fe_2O_3\ (hematite) = CO_2\ (g) + 2Fe_3O_4\ (magnetite) \quad (2)$$

$$CH_4 + 12Fe_2O_3\ (hematite) = CO_2\ (g) + 8Fe_3O_4\ (magnetite) + 2H_2O\ (g) \quad (3)$$

Nitrogen is a minor constituent of the crust and (in decreasing order of importance) is present as fixed (i.e., chemically combined) nitrogen and organic nitrogen in sediments, fixed nitrogen in igneous rocks, elemental N dissolved in the silicates in igneous rocks, fixed nitrogen in coal, and nitrate minerals (e.g., soda niter $NaNO_3$) in caliche deposits (Lodders & Fegley 2010). Upon heating, organic nitrogen and nitrogen fixed as ammonium compounds will be oxidized to $N_2$, and oxidized nitrogen such as nitrates will thermally decompose to $N_2$ and $O_2$.

Crustal sulfur occurs mainly as pyrite ($FeS_2$) and gypsum ($CaSO_4 \cdot 2H_2O$). With increasing temperature, gypsum decomposes to anhydrite ($CaSO_4$), which in turn thermally decomposes and releases $SO_2$ (g). Pyrite is oxidized by reaction with steam forming $SO_2$ and iron oxides. The thermal stability of pyrite and anhydrite accounts for the relatively high temperatures required for $SO_2$ formation in our calculations.

Chlorine and fluorine are mainly present as halide minerals in the crust. Fluorite ($CaF_2$) is the major F-bearing mineral while halite (NaCl), sylvite (KCl), and carnallite ($KMgCl_3 \cdot 6H_2O$) are the major Cl-bearing minerals. Upon heating these minerals react with steam giving HF and HCl. Thus, to first approximation the atmospheric composition resulting from heating up the continental crust to temperatures of 1000 – 1500 K is easy to understand. Phosphorus is mainly present in the crust as phosphate minerals, such as apatite $Ca_5(PO_4)_3(F,OH,Cl)$. Phosphorus oxide gases are produced by thermal decomposition of phosphates heated to high temperatures.

The major features of atmospheric chemistry at higher temperature are also straightforward. Thermal dissociation of $CO_2$, $H_2O$, and $SO_2$ produces CO, OH, SO, O, and H via reactions such as

$$CO_2\ (g) = CO\ (g) + O\ (g) \tag{4}$$

$$H_2O\ (g) = OH\ (g) + H\ (g) \tag{5}$$

$$SO_2\ (g) = SO\ (g) + O\ (g) \tag{6}$$

Knudsen effusion mass spectroscopic (KEMS) studies of the vaporization of silica show that the two major gases are SiO and $O_2$ and that monatomic O is produced by thermal dissociation of $O_2$ (Nagai et al. 1973)

$$2SiO_2\ (silica) = 2SiO\ (g) + O_2\ (g) \tag{7}$$

$$O_2\ (g) = 2O\ (g). \tag{8}$$

KEMS studies of the vaporization of other silicates show that Mg, Ca, SiO, Na, K, $O_2$, and O are formed (see references in Schaefer & Fegley 2004). Subsequent gas phase reactions lead to formation of other high temperature gases containing these elements.

The bulk silicate Earth contains many of the same host phases for C, N, S, Cl, and F so it is not surprising that the predicted atmospheric compositions for the continental crust and BSE are similar. The major difference is probably that the BSE, which is 99.4% mantle by mass, is more reduced than the Earth's crust. Ferrous iron ($Fe^{2+}$) is much more abundant than ferric iron ($Fe^{3+}$) in mantle minerals. The dominance of ferric iron in the crust is due to oxidation of the crust over time by photosynthetically produced oxygen Most of the $O_2$ produced photosynthetically was used to produce ferric iron and sulfate minerals in the crust and only ~ 4% of the total made resides in Earth's atmosphere (Warneck 1989). Earth's crust would be much less oxidized in the absence of life on Earth and the gases resulting from vaporization of crust on a lifeless planet will probably be more reduced.

*4.2 Application to Exoplanets*

The calculations discussed above can be used to predict first order atmospheric compositions for terrestrial planets in a variety of situations and stages of evolution. In particular, we seek to describe the possible ranges of atmospheric composition for an Earth-like planet that is devoid of life and subjected to different stellar environments. Below, we discuss application of our calculations to exoplanet atmospheres, and show two examples of vertical atmospheric profiles.

Figure 9 shows temperature-pressure profiles for the two examples discussed below. The first example is the planet GJ 1214 b, which, although far less dense than the Earth, remains to date the only super-Earth for which atmospheric observations have been made. We use a temperature-pressure profile taken from Miller-Ricci & Fortney (2010) for a hypothetical composition of 50% water vapor and 50% $CO_2$. Their profile was calculated using a 1D radiative-convective climate model and a grey atmosphere approximation.

Our second example is for the very hot exoplanets such as CoRoT-7 b and Kepler-10 b, which are far more Earth-like in their densities. No radiative-convective models have been published for these planets to date that we are aware of. Such a model is beyond the scope of this paper. However, in the interest of providing a first-order approximation, we calculated a simple dry adiabatic lapse rate:

$$\frac{\partial T}{\partial z} = -\frac{\bar{\mu} g}{\bar{C}_p} \tag{9}$$

where $\bar{\mu}$ is the average molecular weight of the gas, $g$ is the gravitational acceleration (35.2 m $s^{-2}$ for CoRoT-7 b), and $\bar{C}_p$ is the average molar heat capacity of the gas. Pressure was determined by assuming hydrostatic equilibrium. The molecular weight and heat capacity were determined from our calculations discussed in §3.3 for our chosen surface conditions. We chose the surface temperature of the substellar point on CoRoT-7b (2500 K, Leger et al. 2011) and a surface pressure for this temperature based on our previous model ($10^{-2}$ bars, Schaefer & Fegley, 2009). Using these temperature-pressure profiles, we then calculated the atmospheric chemistry as a function of height. We discuss these results below.

*4.2.1 Cool Super-Earths*

As mentioned in §1, the mass-radius relationship of GJ 1214 b leads to significant degeneracies in its possible bulk composition. Two main atmospheric compositions have been suggested: (1) H/He envelope, or (2) a water, or "ice", envelope. Nettlemann et al. (2010) give possible H/He mass fractions of 1.3-6%, but cannot constrain the possible water content of this planet to better than 0-97%. Rogers & Seager (2010) find H/He mass fractions of 0.01 – 5%, or $H_2O$ mass fractions > 47%. They also do not rule out an outgassed atmosphere, with 0.3-1.2% by mass of pure $H_2$ necessary for an Earth-like interior structure. Heavier atmospheric gases would decrease the amount of $H_2$ by mass needed, while increasing the total atmospheric mass fraction. Miller-Ricci Kempton et al. (2012) have produced detailed photochemical models for solar and enriched solar composition atmospheres for GJ 1214 b, but so far no detailed calculations have been done to describe a possible outgassed atmosphere.

To illustrate the application of our model, we apply our calculations to determine what atmospheric constituents we would expect if GJ 1214 b had a silicate composition similar to the Earth. Although unlikely, the data available for GJ 1214 b provide a good test case for application of our model. The size and composition of the core will have little to no affect on the outgassing of silicate material, so we do not attempt to constrain it. Figure 10 shows the application of our results to GJ 1214 b. We use the temperature-pressure profile calculated by Miller-Ricci & Fortney (2010) for a 50% $H_2O$ – 50% $CO_2$ atmosphere. We then take the atmospheric composition at the surface temperature (1330 K, 100 bars) and equilibrate it throughout the atmosphere. We find that clouds of alkali salts (NaCl, KCl, RbCl, LiF, etc.) condense at about the 1-10 bar level. These elements are then removed from the atmospheric composition, so the upper atmosphere is depleted in these elements. This, coupled with a rapid temperature drop between the 1 to 0.01 bar levels causes the steep drop off in gas abundances seen in Figure 10 at these pressures. Abundances of other gases remain constant with altitude. (Note that we did not include photochemical effects in these calculations.)

We calculated the scale height, which is given by:

$$H = \frac{RT}{\bar{\mu} g} \tag{10}$$

where $R$ is the ideal gas constant, $g$ is 8.89 m s$^{-2}$ for GJ 1214 b, and $\bar{\mu}$ is the mean molecular weight of the atmosphere. The calculated scale heights for these atmospheres are ~42 and 50 km for the continental crust and the BSE, respectively. In comparison, a pure $H_2$ atmosphere would have a scale height of ~600 km, and a pure $H_2O$ atmosphere would have a scale height of 69.1 km. The model scale heights are therefore too small to account for the deeper Ks-band observations of Croll et al. (2011). However, results for the BSE indicate that it is possible to get substantial amounts of $H_2$ from planetary outgassing, particularly with variations in the total O abundance, and the calculations presented in this paper could be useful in developing comparative spectral models such as those of Miler-Ricci & Fortney (2010).

*4.2.2 Hot Super-Earths*

CoRoT-7 b was the first known planet with a radius close to that of the Earth. The mass of CoRoT-7 b has been the subject of debate (Queloz et al. 2009; Bruntt et al. 2010; Hatzes et al. 2010, 2011; Ferraz-Mello et al. 2010; Pont et al. 2010; Boisse et al. 2011). The uncertain mass has lead to diverse models for the planet's bulk composition ranging from the remnant of a gas giant (Valencia et al. 2010; Jackson et al. 2010) to a volatile-depleted lava world (Léger et al. 2011). The Kepler team discovered a second planet, Kepler-10 b, which has very similar stellar, orbital, and planetary properties to CoRoT-7 b (Batalha et al. 2011). Additionally, Kepler-9 d, whose mass has not yet been directly determined, is estimated to have a similar density based on mass upper limits from transit timing variations (Holman et al. 2010). These planets could be explained with an interior structure similar to that of Mercury (i.e., a large iron/silicate mass ratio). More than likely, these planets migrated to their current orbital period, and may once have had a significant volatile component that was lost over time.

Little is known about the atmospheres of these planets. In our initial models of the atmosphere of CoRoT-7b, we assumed complete loss of all volatile elements (H, C, N, S, etc.) (Schaefer & Fegley 2009; Léger et al. 2011). However, detailed models for the loss of these elements have not yet been done. Valencia et al. (2010) found that the planet could have lost significant mass over time, and that the mass loss rate did not appreciably depend upon the composition or phase of the material being vaporized. However, it has been shown by Tian (2009) that atmospheres with larger molecular weights should be more resistant to evaporative loss. Mass loss should preferentially remove lighter material (H, C, N, etc.) and further increase the mean molecular weight of the atmosphere over time. Therefore, over the planet's lifetime, it may well evolve from something similar to the Earth (Fig. 1 & 2), to volatile-depleted, as described by Schaefer & Fegley (2009). Figures 1 and 2 show that if volatiles remain on this planet, then the signature of evaporating silicates could be obscured by large amounts of steam, $CO_2$, and possibly $SO_2$, for surface temperatures of 1800 – 2600 K for atmospheric pressures greater than ~$10^{-2}$ bars. At lower pressures, the vaporizing silicate material is much more abundant, and $H_2O$ and $CO_2$ dissociate to atomic forms at much lower temperatures. For atmospheric pressures from $10^{-6}$ to $10^{-2}$, the major gases between 1800 – 2600 K are Na, K, O, SiO, and $O_2$, which is not appreciably different from the compositions predicted for the volatile-free compositions of Schaefer & Fegley (2009).

Figure 11 shows our calculations of atmospheric composition along the vertical temperature-pressure profiles shown in Figure 9 (b & c). For these calculations, we set the surface atmosphere composition equal to the gas phase at 2500 K, $10^{-2}$ bars. We took only those gases with mole fractions > $10^{-3}$ for the sake of simplicity, so the compositions of the two atmospheres differ in elemental make-up. The continental crust atmosphere at these conditions contains significantly more O and Si, and less H, Na, Ti, and Fe than the BSE. Also under these conditions, the BSE has significant amounts of N and Cl; by contrast, these elements have abundances < $10^{-3}$ for the continental crust, and so are not included in the vertical calculations presented in this section.

For the vertical profile calculations, the gas is allowed to re-equilibrate at each step upward along the adiabatic profile. Condensation is allowed, in order to simulate cloud formation. Condensed material is removed from the bulk composition so that layers above a cloud are depleted of condensable elements. Photochemistry and other non-equilibrium effects are not included. Gas chemistry is shown in Figure 11 as mole fraction versus total pressure (temperature profiles are different for the left and right panels). We show only gases > $10^{-2}$ mole fraction for legibility. In general, sharp decreases in abundance, when reading from bottom to top, indicate condensation.

The gas chemistry shows that most of the heavy elements (Fe, Si, Ti) condense within a few kilometers of the surface. Lighter elements such as Na, K, C, S, N, H, and O persist into the upper atmosphere. For both compositions, Na condenses out around (1-2)$\times 10^{-4}$ bars, primarily as $Na_2CO_3$ for the continental crust and $Na_2O$ + NaOH for the BSE, whereas K persists up to ~ (4-5)$\times 10^{-5}$ bars, and condenses primarily as KOH (+KCl for the BSE). Carbon and sulfur condense around $5\times 10^{-4}$ bars as $Na_2CO_3$ and $Na_2SO_4$ (+ $K_2SO_4$ for the BSE), respectively. Hydrogen condenses as KOH for the continental crust, and is completely removed from the atmosphere at ~ $4\times10^{-5}$ bars. For the BSE, hydrogen condenses as both NaOH and KOH, but is not completely removed from the atmosphere due to differences in elemental abundance. At higher altitudes the remaining gas is only made up of $O_2$, in the case of the continental crust, and $O_2 + H_2O + N_2$ for the BSE. For the BSE, $H_2O$ and KCl show a sharp decrease in abundance at ~$6\times10^{-4}$ bars, and an increase at lower pressures. This dip is related to the increase in NaOH and NaCl gas abundances. As Na starts to condense at higher altitudes, the abundances of the Na gases drops, returning H and Cl to $H_2O$ and KCl gases, respectively.

The results of these calculations suggest that the upper atmospheres of hot rocky planets may be significantly different from the lower atmospheres. This may explain the lack of detection of exospheric Na and Ca by Guenther et al. (2011). However, these calculations are first-order approximations and should be taken with a grain of salt. A radiative-convective climate model which accounts for tidal-locking effects may give a significantly different temperature-pressure profile. Additionally, the simple cloud condensation model used here may not be accurate; given the low atmospheric pressures involved, condensation may be limited to simpler molecules than considered here, and may depend on the availability of nucleation sites. However, we note that in comparison to volatile-free atmospheres (Schaefer & Fegley 2009), we find that the condensation altitudes of the alkali metals are higher when including volatiles (H, C, N). For instance, in our volatile-free model we found that Na condensed as $Na_2O$ below altitudes of ~28 km, in contrast to altitudes found here of ~44 km.

*4.3 Photochemistry*

Photochemical effects are a major source of disequilibrium in the atmosphere of a non-biogenic super-Earth, particularly for $a < 1$ AU. Full quantitative photochemical models require knowledge of the planet's orbital location and the stellar properties. Since our goal is not to model a particular planet but to provide a tool for future modeling, a full photochemical model is beyond the scope of this paper. We have previously discussed photochemical processes for $CO_2$, CO, $SO_2$, and $H_2O$ in the atmospheres of Venus-like exoplanets (Schaefer & Fegley 2011). Long term photochemical destruction of the atmospheres generated by outgassing of an Earth-like planet (e.g., dominated by $H_2O$ and $CO_2$) would lead to loss of hydrogen due to photodissociation of water vapor and to conversion of $CO_2$ into CO in the upper atmosphere, similar to Venus. As on Venus and Mars, catalytic cycles involving OH, NO, Cl, or F would likely be necessary to re-form $CO_2$ from CO. In the presence of abundant water vapor, photolyis of $SO_2$ can lead to the formation of $H_2SO_4$ clouds via the net reaction:

$$SO_2 + H_2O + \frac{1}{2} O_2 = H_2SO_4 \qquad (11)$$

as found on Venus. Given that $SO_2$ is significantly less abundant than $H_2O$ in the atmospheres considered here, it is possible that production of $H_2SO_4$ clouds would severely deplete the upper atmosphere in $SO_2$, while allowing some water to remain in the atmosphere above the clouds. However, $H_2SO_4$ clouds will dry out the upper atmosphere via formation of hydrated sulfuric acid (1 or more $H_2O$ molecules per $H_2SO_4$).

We estimated lower limits to the photochemical lifetimes of several of the important gases in our models and give the results in Table 4. These calculations were done using photodissociation rates calculated for the top of Earth's atmosphere at 1 AU from Levine (1985) and Matveev (1986). We then scaled the solar flux as a function of radial distance. In the table, we show photochemical lifetimes at the orbital distances of Earth (1 AU), Kepler-11 f (0.25 AU), and the average value for CoRoT 7 b and Kepler-10 b (0.017 AU). The Kepler and CoRoT surveys are targeting planets that orbit stars similar to the Sun, making the use of solar insolation a reasonable approximation. Photochemical lifetimes for the major gases are on the order of seconds to minutes for CoRoT-7b and Kepler-10 b. The longest-lived compound is CO with a lifetime of ~7.5 minutes, and the shortest is SiO, with an estimated lifetime of only a thousandth of a second. Kepler-11 f is currently the super-Earth furthest from its parent star, and has slightly longer photochemical lifetimes on the order of seconds for SiO and $H_2S$, and up to a day for CO.

While photochemical lifetimes for planets discovered by Kepler and CoRoT can be estimated using solar values, another major transiting survey – MEarth – is searching for planets around M stars. The first super-Earth discovered by this survey, GJ 1214 b, has the shortest orbital period of any of the currently known super-Earths, yet has one of the lowest equilibrium temperatures. This is because the luminosity of GJ 1214 is roughly 100 times lower than that of the Sun (Segura et al. 2005). If we scale the photochemical lifetimes in Table 4 based on luminosity ratios of an M star to the Sun, we coincidentally get photochemical lifetimes nearly identical to those of Kepler-11 b. However, simply scaling the flux is not a very good approximation of the true photochemical rates at GJ 1214 b, since M stars typically have larger UV fluxes at wavelengths shortward of 200 nm than Sun-like stars. Therefore, the photochemistry of GJ 1214 b is likely to be different than that of the other planets considered here.

*4.4 Spectroscopically Observable Gases*

Spectroscopic models for super-Earth exoplanets have focused on potential biomarkers ($CH_4$, $NH_3$, $O_3$, $N_2O$) (e.g., Kaltenegger et al. 2007, Segura et al. 2003, 2005, 2010). $CH_4$ and $NH_3$ are very abundant at low temperatures in the atmosphere of the BSE (Figure 2). In the continental crust, however, $CH_4$ is much less abundant, with abundances greater than 1 ppb at temperatures above 600 K for high pressures (> 10 bars), compared to the BSE, which has abundances greater than 1 ppb for T < 1500 K at 100 bars. The peak abundance for $CH_4$ in the continental crust atmosphere is 0.68 ppm at 100 bars and 500 K, compared to ~95% in the BSE. Similarly, ammonia has abundances greater than 1 ppb below 900 K at 100 bars with a peak abundance of 2.8 ppm at 100 bars and 600 K in the continental crust, compared to 0.6% in the BSE. However, the abundances of these gases are dependent on the total oxygen abundance, which is difficult to constrain. Our calculations with variable O abundance showed that minor increases in the total oxygen abundance of just 0.6% in the BSE would prevent the formation of $CH_4$ and $NH_3$.

Ozone and $N_2O$ are both molecules that are considered strong indicators of life. However, both gases can be produced through chemical equilibrium in trace amounts under certain conditions. For example, we find abundances of $O_3$ greater than 1 ppb for T > 1800 K, and pressures above ~0.1 bar in the continental crust, with a peak abundance of 8.7 ppm at 100 bars. This abundance is comparable to that of the ozone layer on the Earth. Ozone abundances in the BSE are similar, with $X_{O_3}$ > 1 ppb for temperatures above 2000 K and pressures above $10^{-2}$ bars. The maximum ozone abundance for the BSE is 6.5 ppm at 100 bars. However, ozone may be photochemically produced in the upper atmospheres of hot planets, which are very rich in $O_2$. In the continental crust models, $N_2O$ has abundances greater than 1 ppb at pressures greater than 1 bar and temperatures above 1700 K. It has a peak abundance of 79 ppb, which is smaller than the observed abundance in the Earth's atmosphere (~320 ppb) where it is produced biologically during denitrification. For the BSE, $N_2O$ has abundances greater than 1 ppb at 2400 – 3500 K only at 100 bars, with a peak value of 4.8 ppb.

$SO_2$ is a spectroscopically interesting molecule that is generally not included in terrestrial exoplanet atmospheric models based on the Earth. However, Kaltenegger & Sasselov (2010) have shown that $SO_2$ would be detectable at abundances of a few ppm for wavelengths between 4-40 μm. Our model predicts $SO_2$ abundances of ~3% at temperatures above 1300 K for the continental crust, and 5-20 % between 1700 – 3500 K. This suggests that $SO_2$ should be included when generating models of atmospheric spectra for terrestrial exoplanets.

High temperature evaporating planets will have a different spectroscopic signature than the cooler planets. At high temperatures and prior to significant volatile loss, the atmosphere would still be dominated by $H_2O$ and $CO_2$ up to temperatures of 3000 K for pressures above ~1 bar. NaOH, KOH, NaCl, KCl, Na, K, and SiO could also contribute to the planet's transmission spectrum, particularly at lower pressures, which may be more applicable to an atmosphere produced by planetary vaporization. However, as shown by the calculations in Figures 9 &11, the alkalis may condense at cooler levels of an atmosphere, which would remove them from the upper atmosphere and reduce their contribution to the planetary spectra. Other abundant species that may contribute to the transmission spectrum include CO, OH, and NO at high temperatures. Additionally, photodissociation and photoionization should be rapid in the upper atmospheres of evaporating planets. The exospheres of these planets are likely to be dominated by O, H, N, C, S, Na, K, and their ions.

## 5. SUMMARY

Vaporization of the crust of an Earth-like planet produces atmospheres rich in $H_2O$ and $CO_2$ over broad temperature and pressure ranges. The BSE also produces significant amounts of $SO_2$ at high temperatures and $CH_4$ and $NH_3$ at low temperatures. These gases are spectroscopically important albeit photochemically unstable, and should be considered in future spectroscopic models of outgassed atmospheres. Atmospheres of high temperature evaporating planets at low pressures are not appreciably different from the volatile-free atmospheres calculated by Schaefer & Fegley (2009). Future work will further incorporate kinetic and photochemical processes into our models of hot Earth-like atmospheres, and investigate the interactions of outgassed atmospheres with captured nebular gas. Detailed chemical models of volatile loss over time would greatly enhance our ability to predict the atmospheric composition of hot super-Earths.

This work was supported by the NSF Astronomy Program AST- 0707377 and by NASA Cooperative Agreement NNX09AG69A with the NASA Ames Research Center.

**Table 1.** Orbital and physical properties of known transiting low-mass planets

| Planet | Mass ($\times M_\oplus$) | Radius ($\times R_\oplus$) | Density (g cm$^{-3}$) | Orbital Radius (AU) | $T_{eq}$ (K) |
|---|---|---|---|---|---|
| Kepler-11 f[1] | 2.30 | 2.61 | 0.71 | 0.25 | 544 |
| GJ 1214 b[2] | 6.55 | 2.68 | 1.88 | 0.014 | 555 |
| HD 97658 b | 6.40 | 2.93 | 1.40 | 0.0797 | 510-720 |
| Kepler-11 e[1] | 8.40 | 4.52 | 0.5 | 0.194 | 617 |
| Kepler-11 d[1] | 6.10 | 3.43 | 0.83 | 0.159 | 692 |
| Kepler-11 b[1] | 4.30 | 1.97 | 3.10 | 0.091 | 900 |
| Kepler-20 b | 8.70 | 1.91 | 6.5 | 0.04537 | 1014 |
| Kepler-18 b | 6.90 | 2.00 | 4.9 | 0.0447 | 1179 |
| CoRoT-7 b[3] | 7.42 | 1.58 | 10.4 | 0.0172 | 1810 |
| Kepler-10 b[4] | 4.54 | 1.42 | 8.74 | 0.01684 | 1833 |
| 55 Cnc e | 7.98[5]/8.63[6] | 2.13[5]/2.00[6] | 4.97[5]/5.9[6] | 0.01564 | 1967 |

[1]Lissauer et al. 2011. [2]Charbonneau et al. 2009. [3]Léger et al. 2009; Hatzes et al. 2011. [4]Batalha et al. 2011. [5]Demory et al 2011. [6]Winn et al. 2011. [7]

**Table 2.** Bulk compositions of vaporizing planets

| Element | Continental Crust[1] (wt%) | Bulk Silicate Earth[2] (wt%) |
|---|---|---|
| O | 47.20 | 44.42 |
| Si | 28.80 | 21.61 |
| Al | 7.96 | 2.12 |
| Fe | 4.32 | 6.27 |
| Ca | 3.85 | 2.46 |
| Na | 2.36 | 0.29 |
| Mg | 2.20 | 22.01 |
| K | 2.14 | 0.02 |
| Ti | 0.401 | 0.12 |
| P | 0.076 | 0.008 |
| Cr | 0.013 | 0.29 |
| Mn | 0.072 | 0.11 |
| H | 0.045 | 0.006 |
| C | 0.199 | 0.006 |
| N | 0.006 | $0.88\times10^{-4}$ |
| S | 0.070 | 0.027 |
| F | 0.053 | 0.002 |
| Cl | 0.047 | 0.004 |
| TOTAL[3] | 99.822 | 99.776 |

[1]Wedepohl (1995). [2] Kargel & Lewis (1993) [3]Totals are less than 100% because Ni is not considered.

**Table 3.** Comparison of the major atmospheric compositions at extreme temperature and pressure conditions for the continental crust and the *bulk silicate Earth (BSE)*.

| | low T<br>1000 K | High T<br>2000 K |
|---|---|---|
| low P<br>$10^{-6}$ bars | 49% $H_2O$, 39% $CO_2$, 4.5% HF<br>*61% $H_2O$, 20% $SO_2$, 12% $CO_2$* | 52% SiO, 25% O, 15% $O_2$<br>*28% Mg, 26% O, 24% SiO* |
| high P<br>$10^{+2}$ bars | 55% $H_2O$, 42% $CO_2$, 1.4% HCl<br>*80% $H_2O$, 16% $CO_2$, 2.3% $H_2$* | 48% $H_2O$, 38% $CO_2$, 4% HF<br>*62% $H_2O$, 20% $SO_2$, 12% $CO_2$* |

**Table 4**. Photochemical lifetimes of important gases at different orbital distances

| | $J_1$ ($s^{-1}$) [a] | $t_{chem}$ (s) | | |
|---|---|---|---|---|
| gas | 1 AU | 1 AU | 0.25 AU | 0.017 AU |
| $H_2O$ | $1.18\times10^{-5}$ | $8.47\times10^{4}$ | $5.29\times10^{3}$ | $2.45\times10^{1}$ |
| $CO_2$ | $2.02\times10^{-6}$ | $4.95\times10^{5}$ | $3.09\times10^{4}$ | $1.43\times10^{2}$ |
| CO | $6.46\times10^{-7}$ | $1.55\times10^{6}$ | $9.68\times10^{4}$ | $4.47\times10^{2}$ |
| $SO_2$ | $2.49\times10^{-4}$ | $4.01\times10^{3}$ | $2.51\times10^{2}$ | $1.16\times10^{0}$ |
| $O_2$ | $4.86\times10^{-6}$ | $2.06\times10^{5}$ | $1.29\times10^{4}$ | $5.94\times10^{1}$ |
| $CH_4$ | $7.79\times10^{-6}$ | $1.28\times10^{5}$ | $8.02\times10^{3}$ | $3.71\times10^{1}$ |
| $NH_3$ | $1.74\times10^{-4}$ | $5.73\times10^{3}$ | $3.58\times10^{2}$ | $1.66\times10^{0}$ |
| $H_2S$ | $3.30\times10^{-3}$ | $3.03\times10^{2}$ | $1.89\times10^{1}$ | $8.76\times10^{-2}$ |
| NO | $3.53\times10^{-6}$ | $2.83\times10^{5}$ | $1.77\times10^{4}$ | $8.19\times10^{1}$ |
| $N_2$ | $1.02\times10^{-6}$ | $9.76\times10^{5}$ | $6.10\times10^{4}$ | $2.82\times10^{2}$ |
| SiO | $4.03\times10^{-3}$ [b] | $2.48\times10^{2}$ | $1.55\times10^{1}$ | $7.17\times10^{-2}$ |

[a] top of the atmosphere, Levine (1985).

[b] estimated based on data from Matveev (1986).

**FIGURE CAPTIONS**

*Figure 1.* Major gas composition as a function of temperature at 100 bars for the continental crust: (a) volatile elements H, C, N, S and (b) rock-forming elements Na, K, Fe, Si, Mg, Ca, Al, and Ti.

*Figure 2.* Major gas composition as a function of temperature at 100 bars for the bulk silicate Earth: (a) volatile elements H, C, N, S and (b) rock-forming elements Na, K, Fe, Si, Mg, Ca, Al, and Ti.

*Figure 3.* Effect of varying hydrogen abundance on gas composition: (a) continental crust, (b) BSE. Panels from top to bottom show 0.1, 1, and 10× the nominal hydrogen abundance. All other elements are kept constant. Calculations are done at 100 bars total pressure.

*Figure 4.* Effect of varying carbon abundance on gas composition: (a) continental crust, (b) BSE. Panels from top to bottom show 0.1, 1, and 10× the nominal carbon abundance. All other elements are kept constant. Calculations are done at 100 bars total pressure.

*Figure 5.* Effect of varying oxygen abundance on gas composition: (a) continental crust, (b) BSE. Panels on top and bottom show the nominal oxygen abundance ± 4 wt%. All other elements are kept constant. Calculations are done at 100 bars total pressure.

*Figure 6.* The gas-to-solid molar ratio as a function of temperature for the selected pressures $10^{-6}$, $10^{-4}$, $10^{-2}$, $10^{0}$, and $10^{+2}$ bars. Results are shown for (a) the continental crust, and (b) the BSE. A small ratio corresponds with large solid fractions, and a large ratio corresponds to a primarily gaseous fraction.

*Figure 7.* Major gas abundances as a function of temperature for different pressures. Solid lines are for the continental crust composition, dashed lines are for the BSE. (a)$H_2O$ (b) $CO_2$ (c) $O_2$ (d) SiO (e) $SO_2$.

*Figure 8.* The most abundant gases for the volatile elements as a function of temperature and pressure. Lines indicate equal mole fraction abundances of gases. Solid lines are for the continental crust and dashed lines are for the BSE. Results for the two compositions are nearly identical. (a) hydrogen (b) carbon (c) nitrogen (d) sulfur.

*Figure 9.* Temperature-pressure profiles for three test cases. Profile *a* was calculated by Miller-Ricci & Fortney (2010) for GJ 1214 b assuming a 50% $H_2O$ + 50% $CO_2$ composition. Profiles *b* and *c* were calculated using a dry adiabatic lapse rate and hydrostatic equilibrium for the continental crust and BSE, respectively, starting from surface conditions appropriate for CoRoT-7b.

*Figure 10.* Atmospheric composition for a planet similar to GJ 1214 b using the temperature pressure profile for an atmosphere of 50% $H_2O$ + 50% $CO_2$ from Miller-Ricci & Fortney (2010) shown in Figure 9 (left: continental crust, right: BSE).

*Figure 11.* Atmospheric composition for a planet similar to CoRoT-7 b using the dry adiabatic temperature pressure profiles shown in Figure 9. Starting compositions were taken for the continental crust (left) and the BSE (right) at 2500 K and $10^{-2}$ bars.

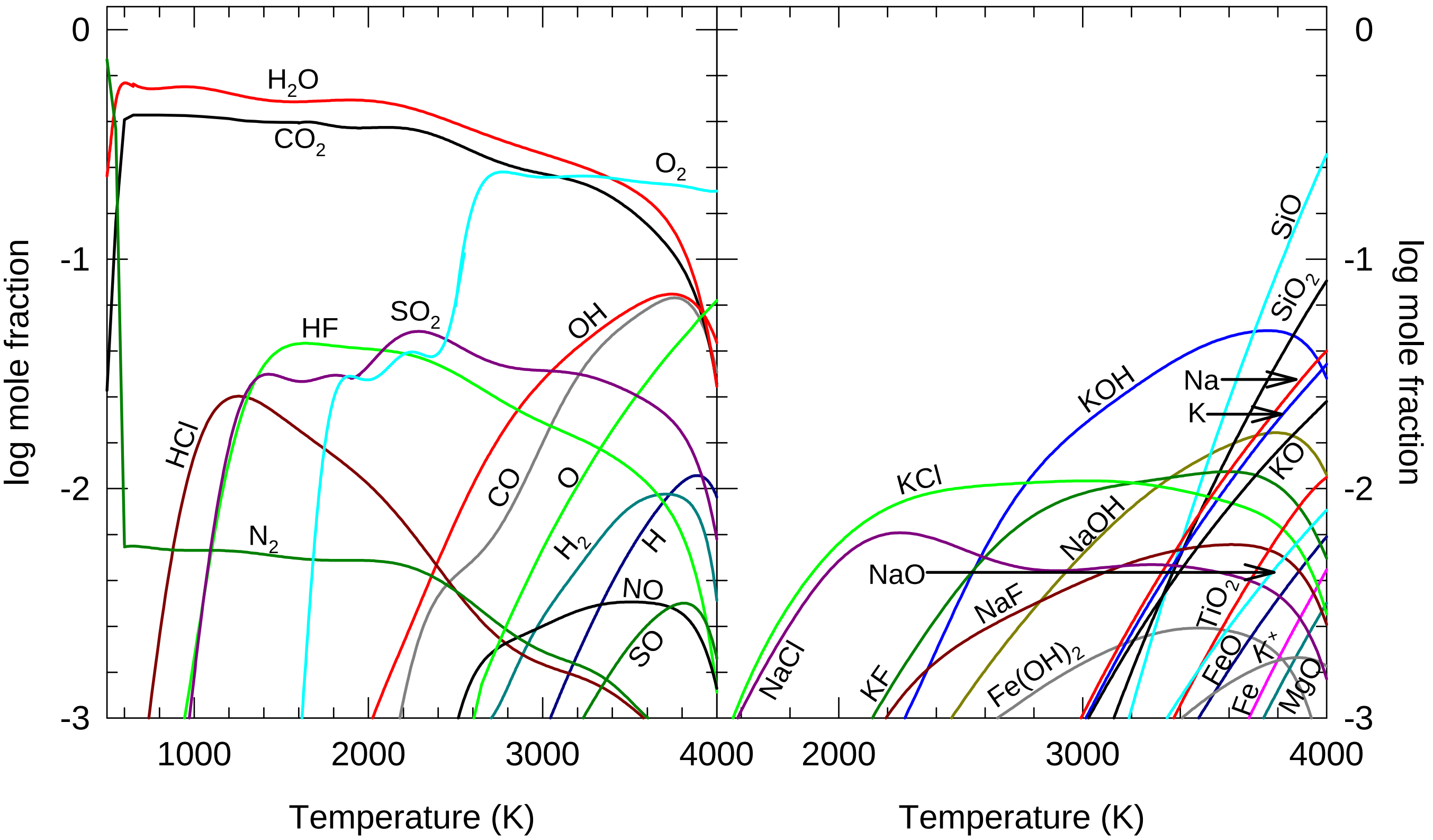

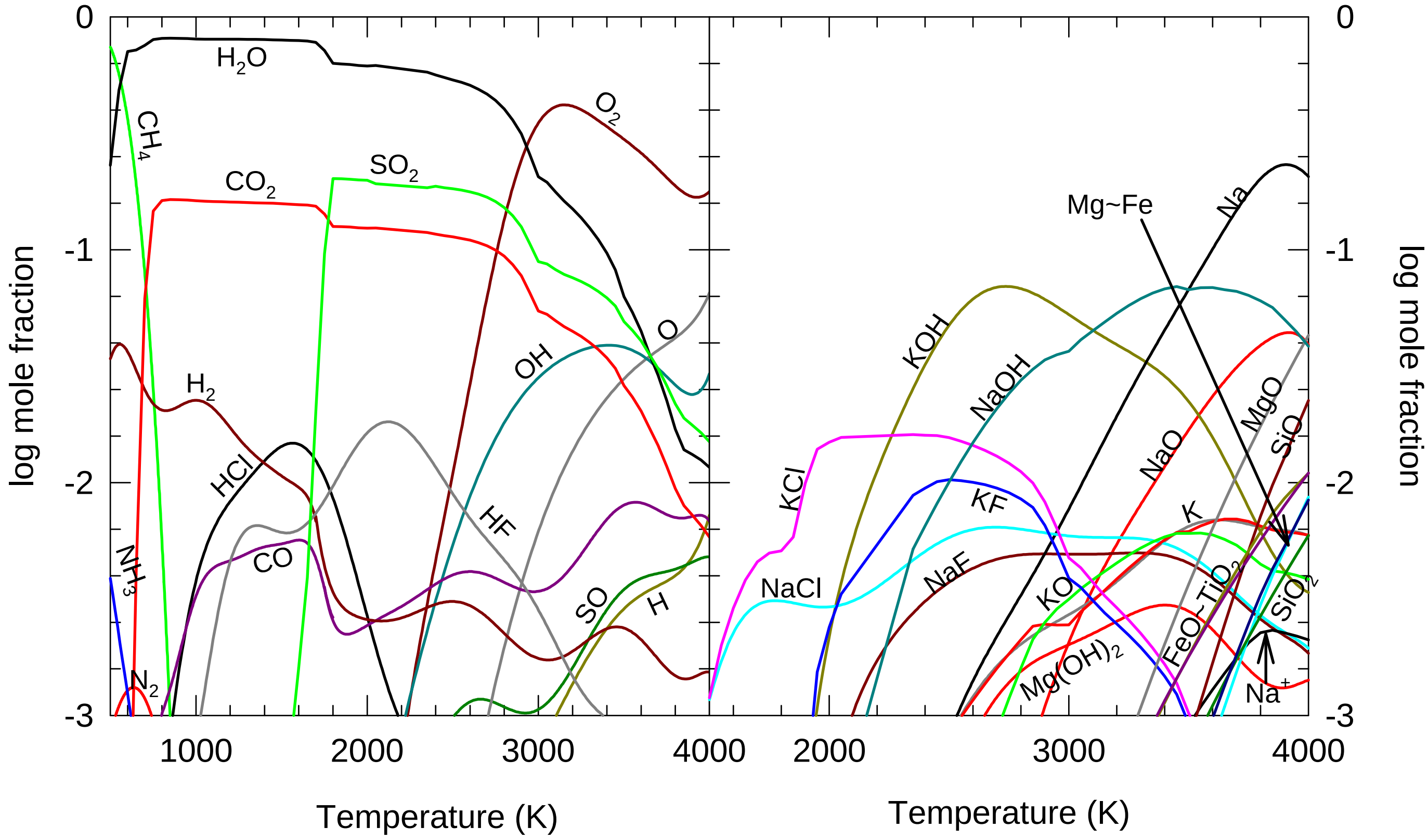

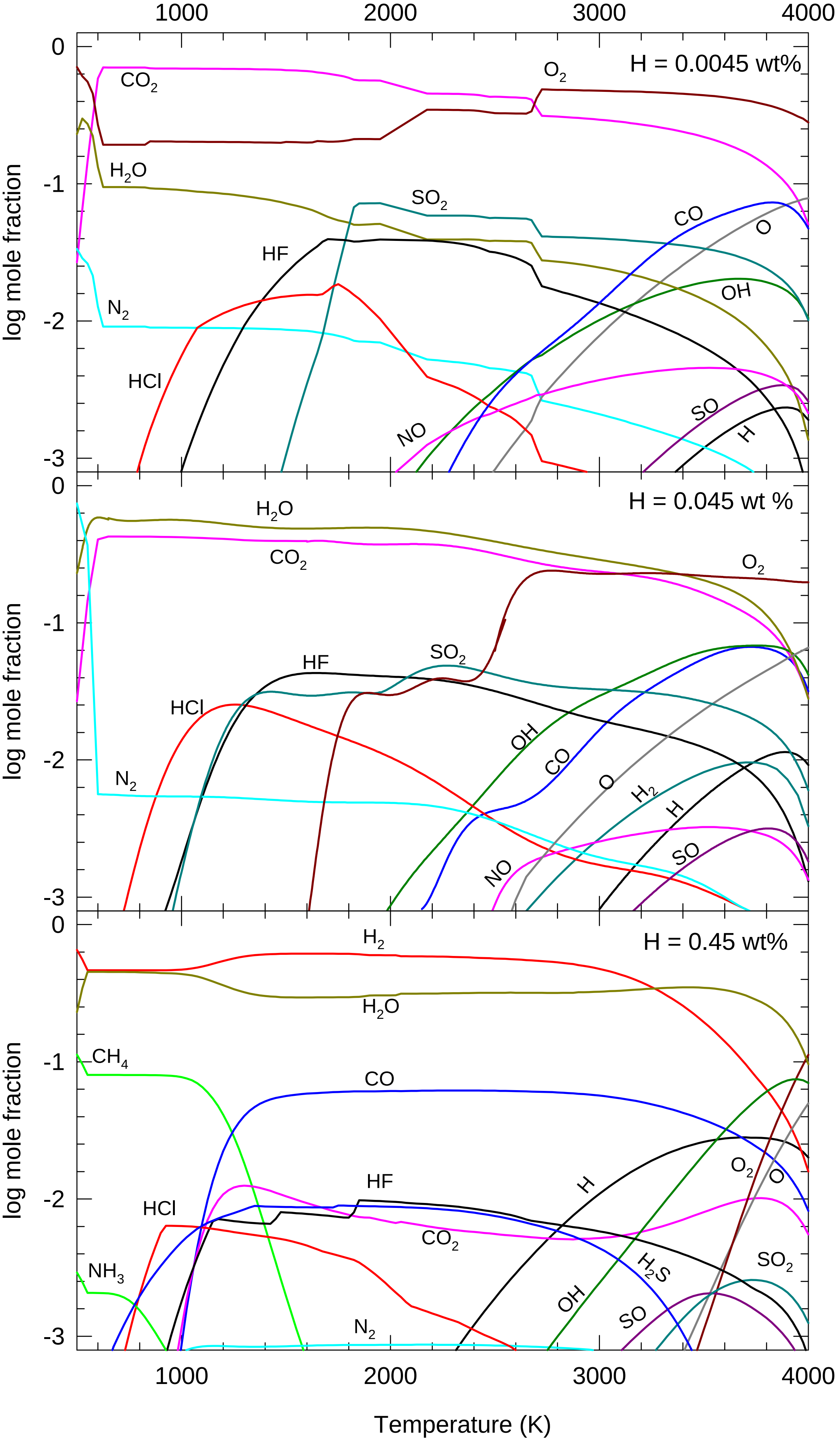

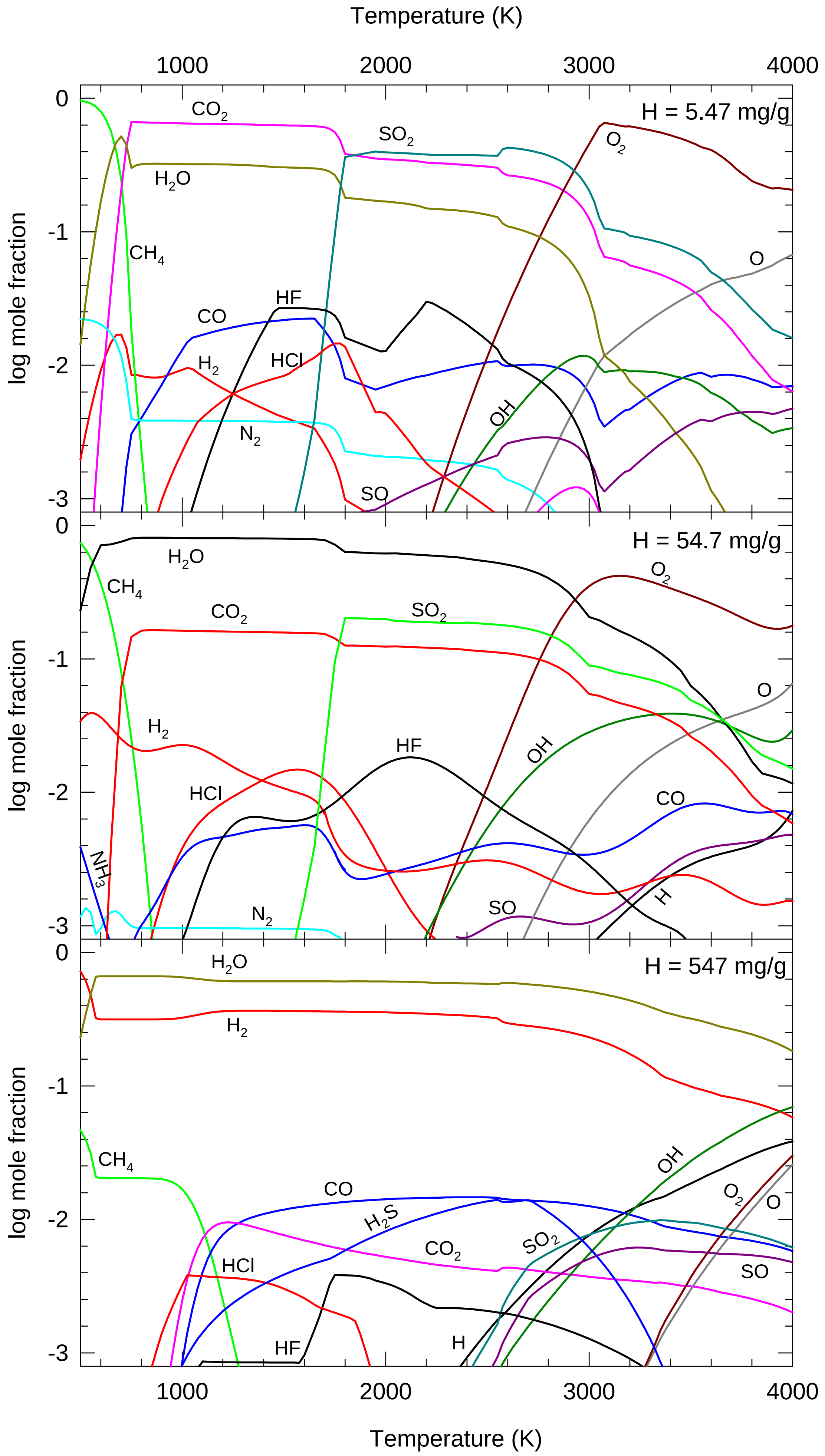

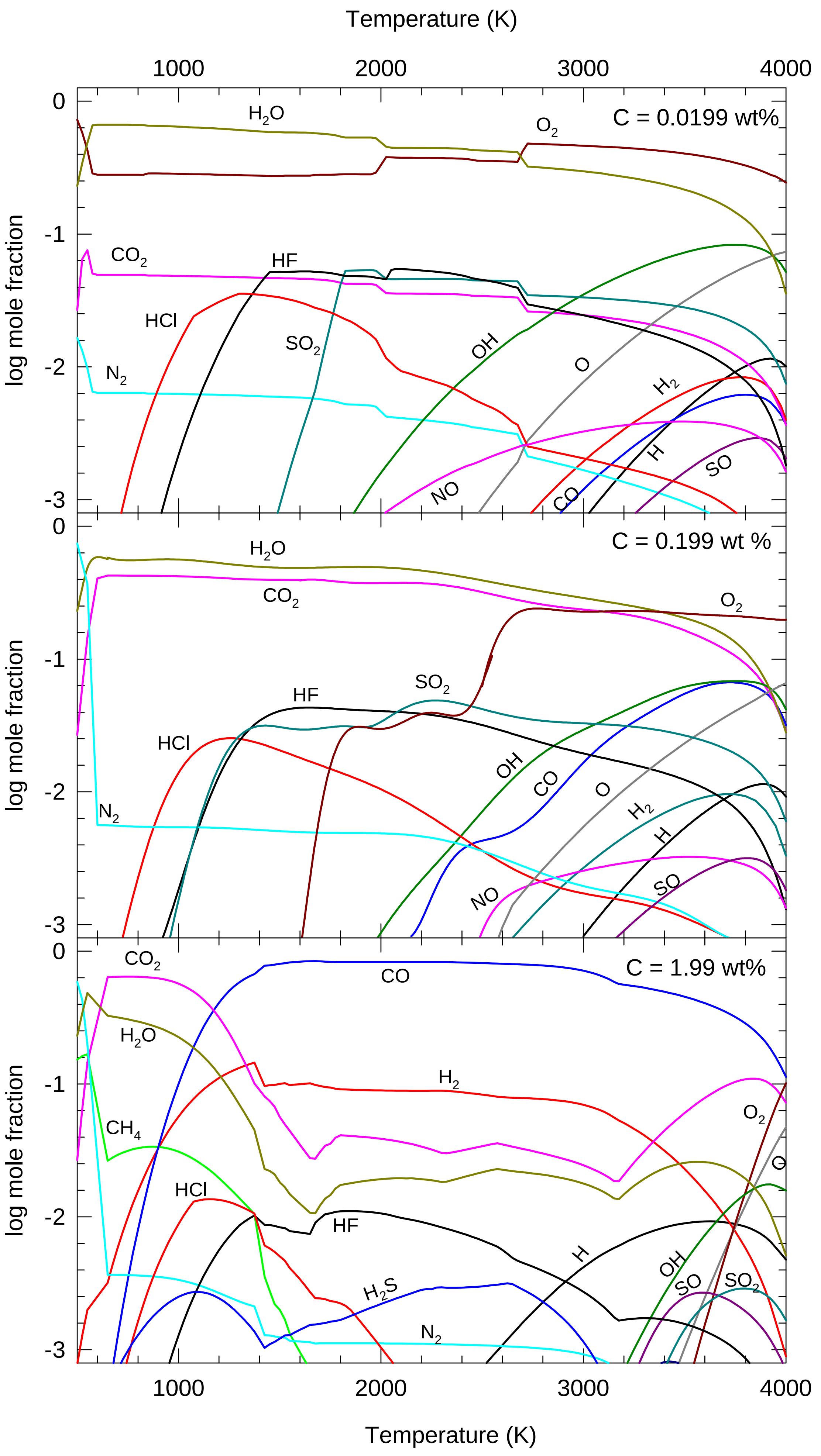

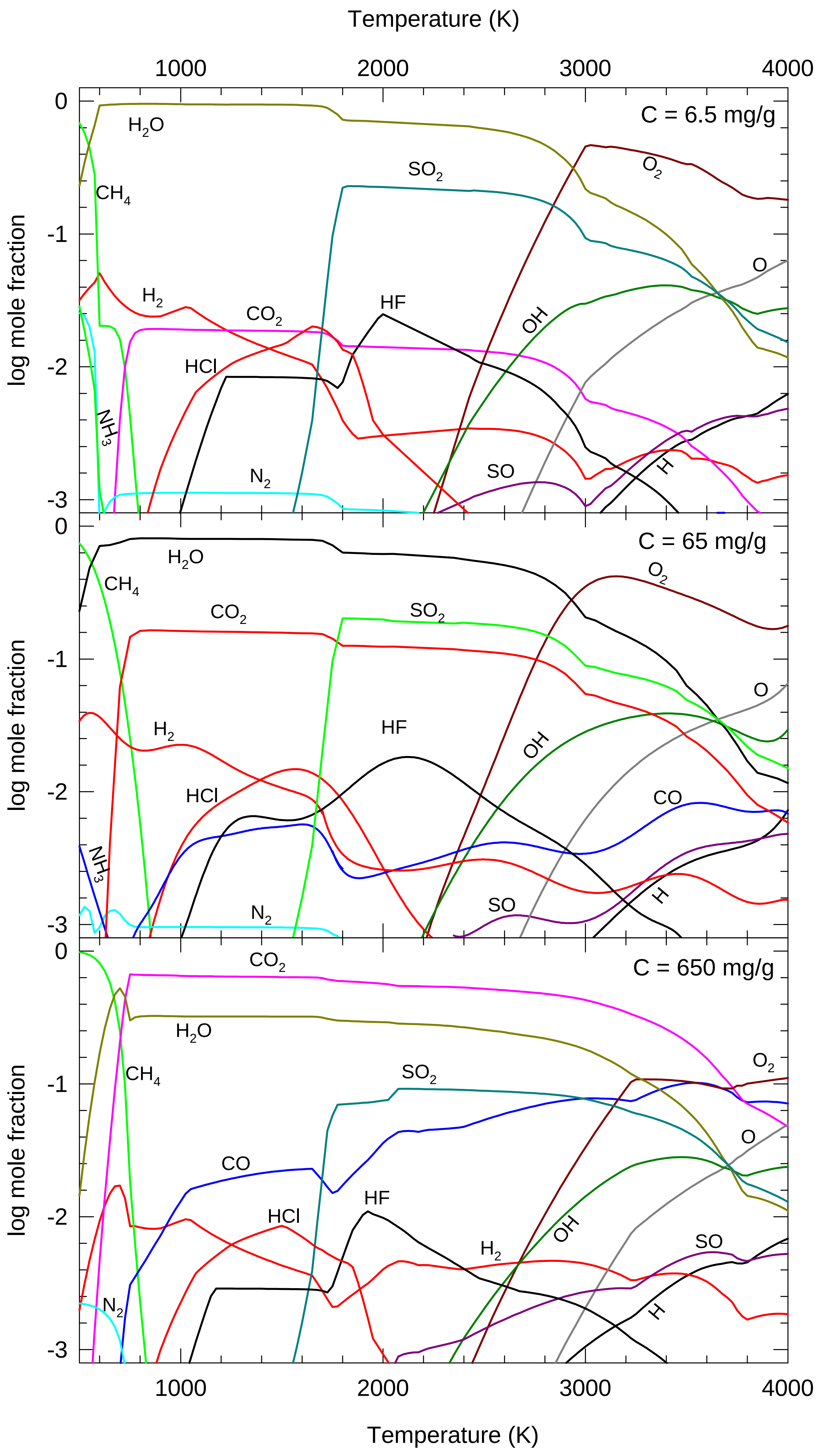

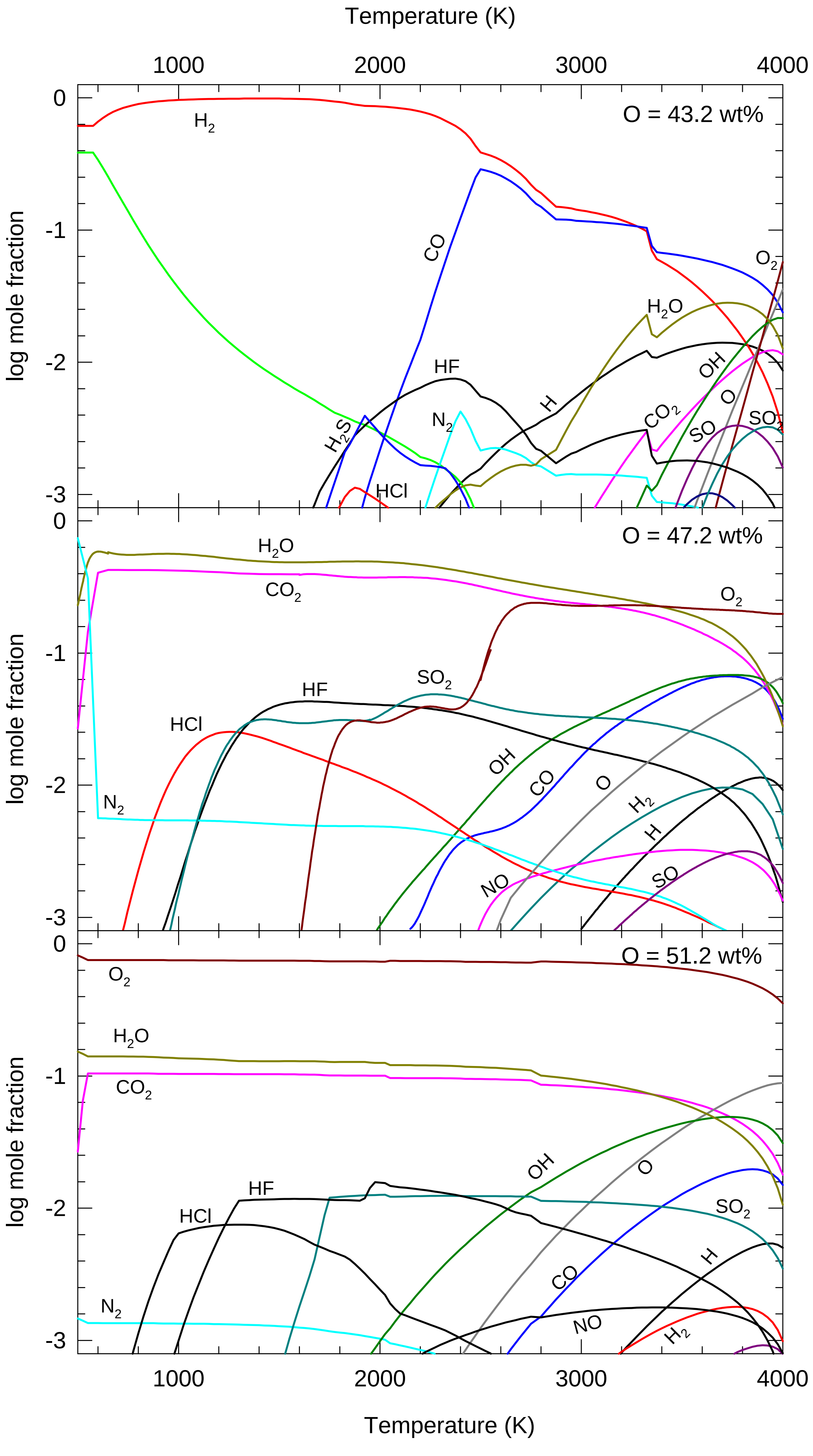

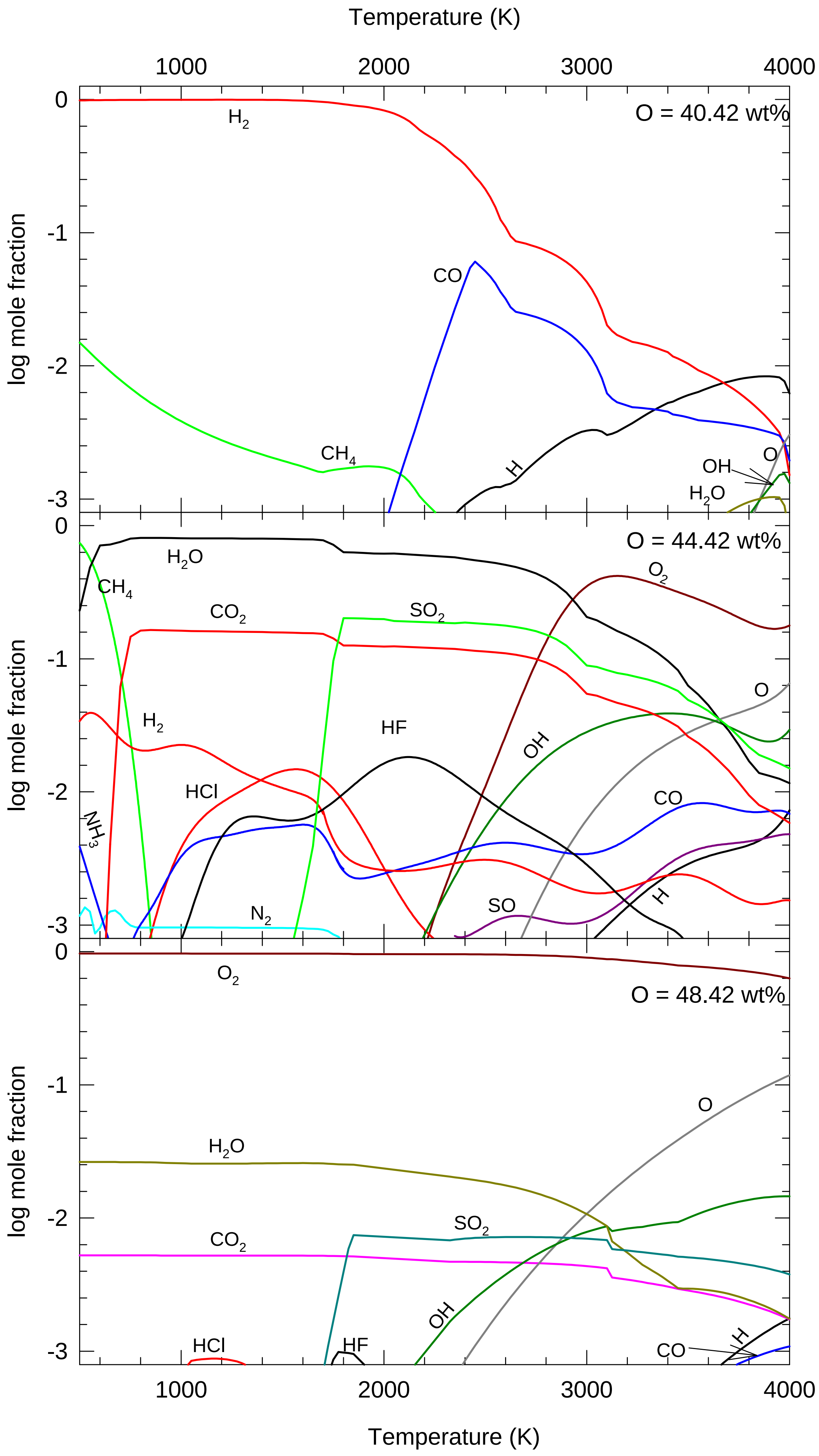

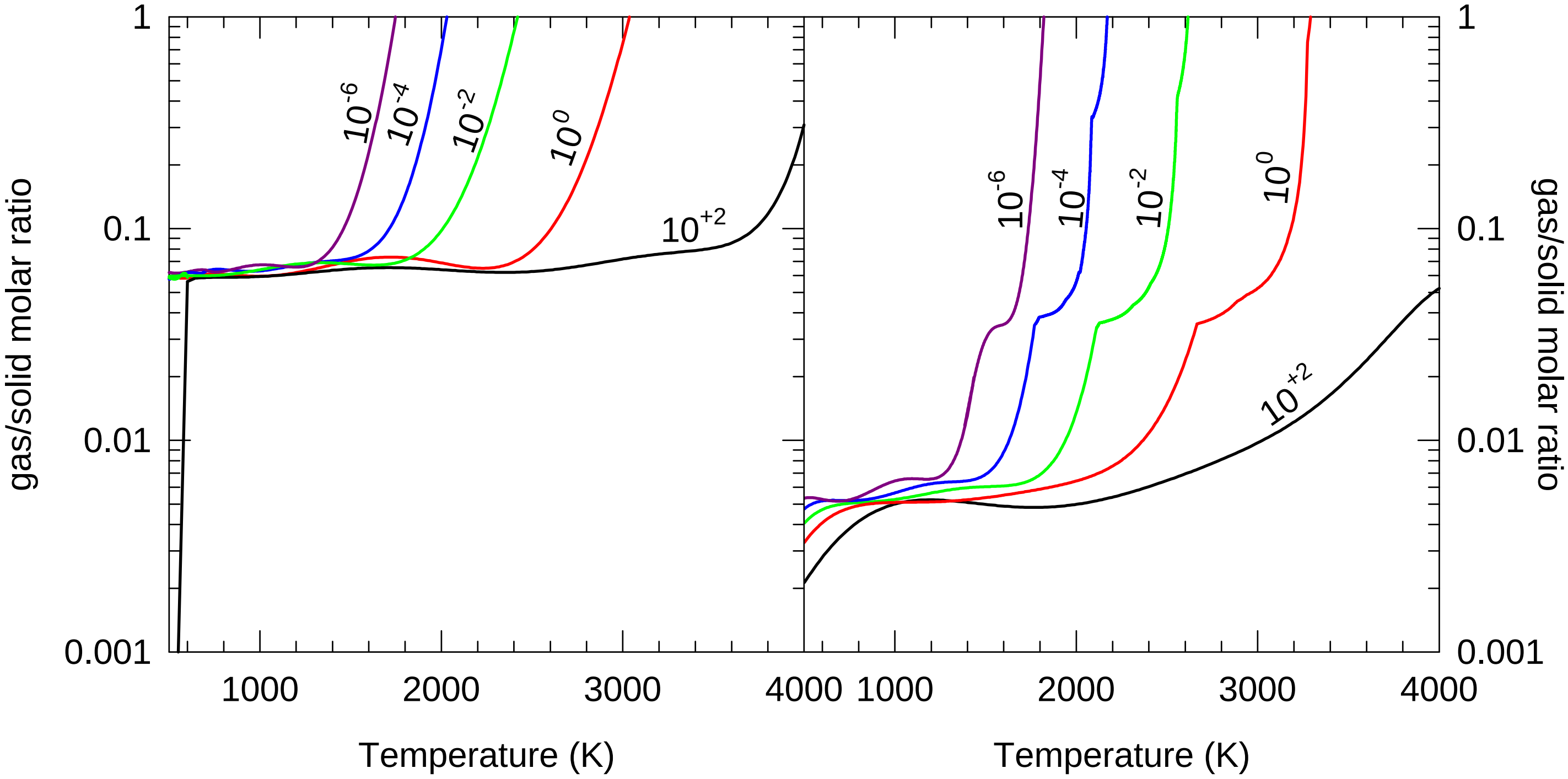

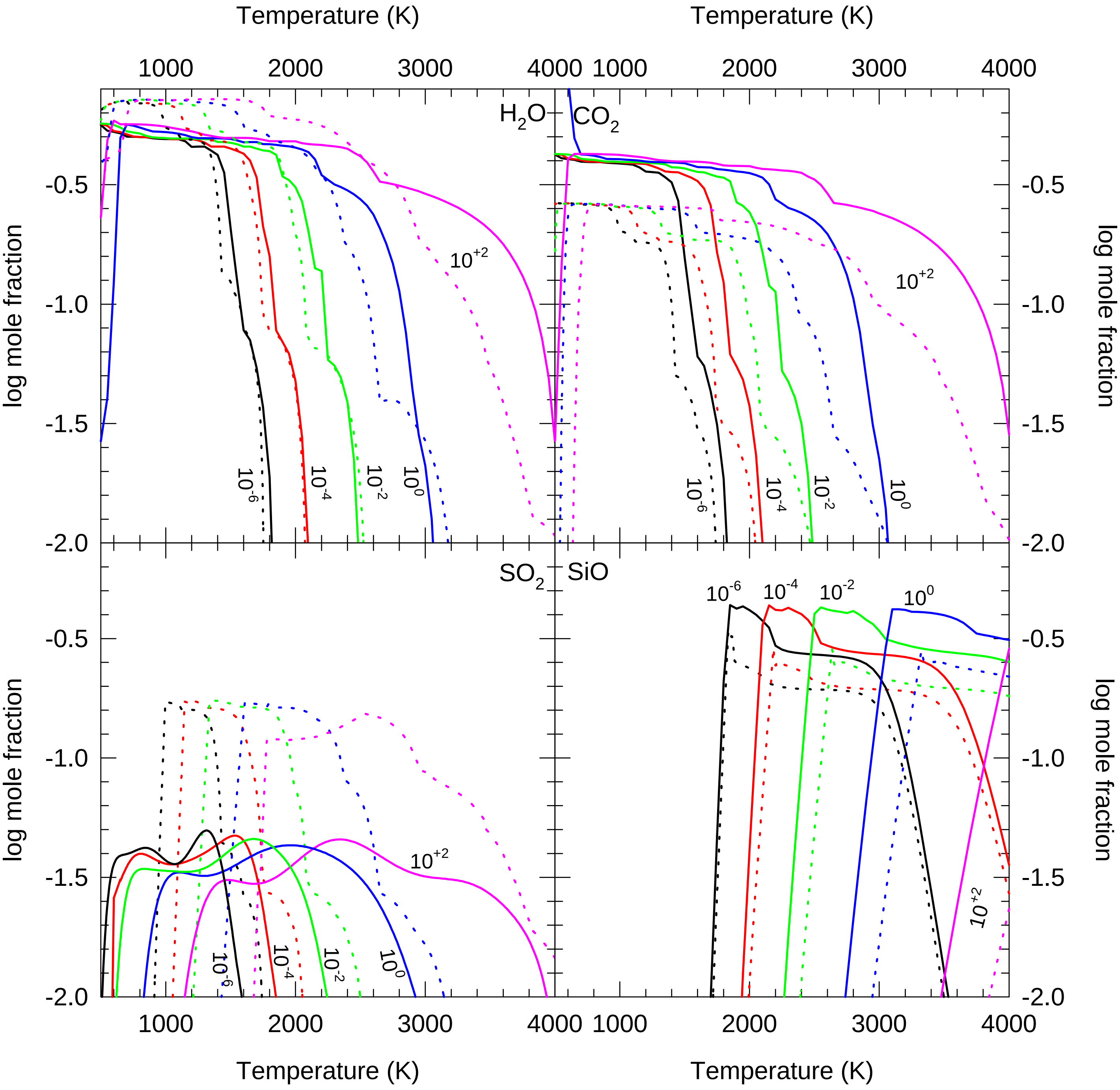

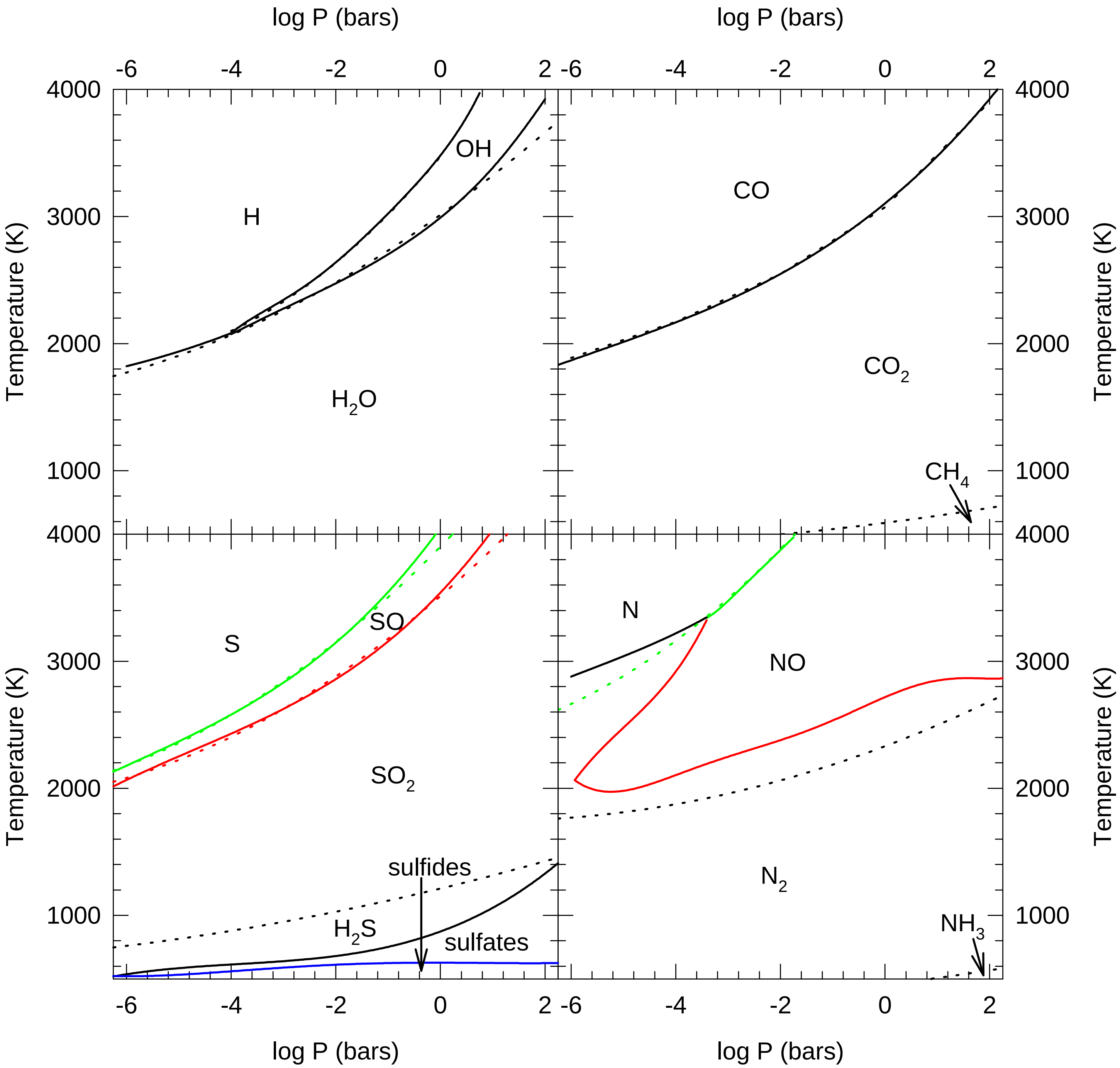

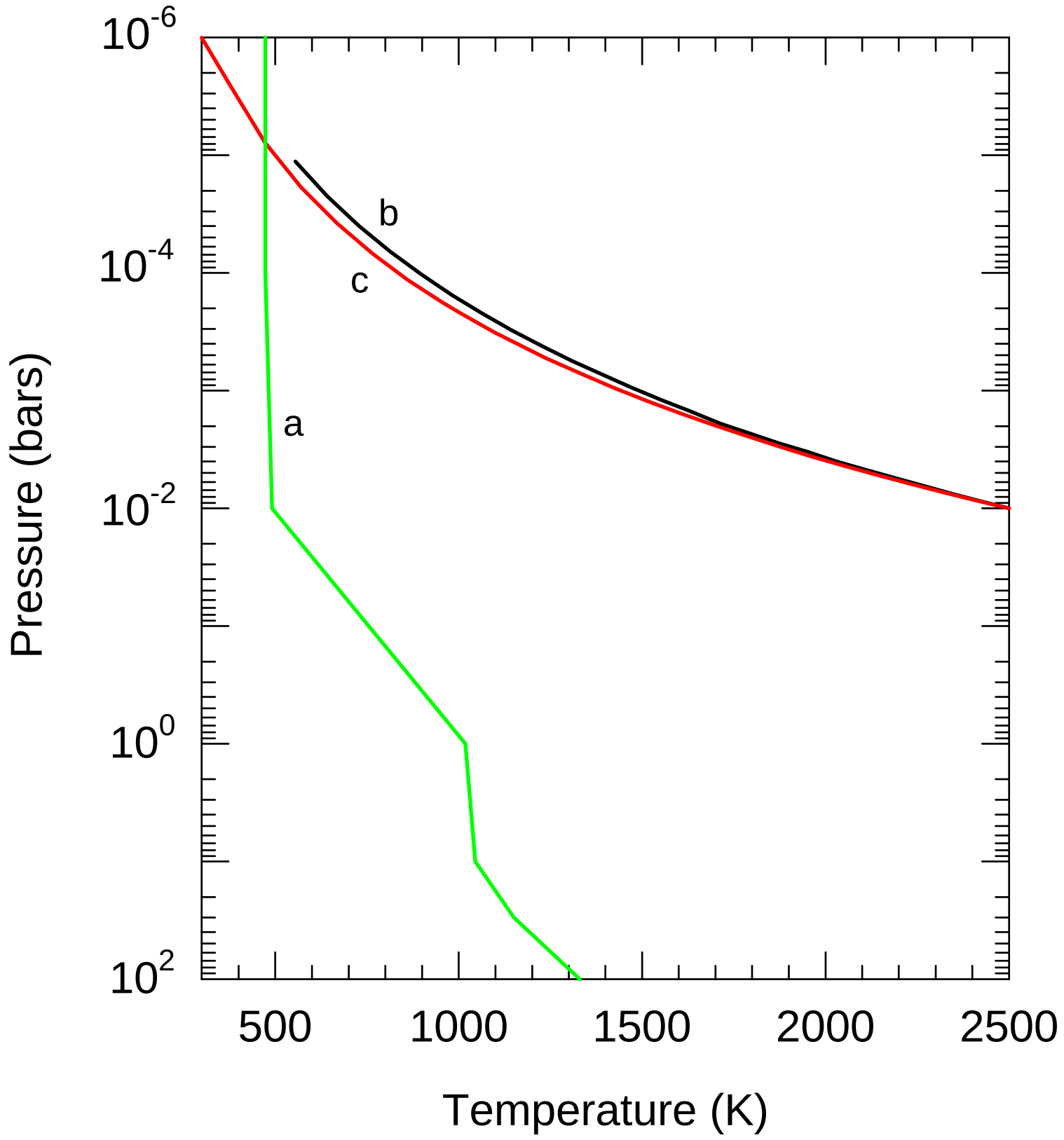

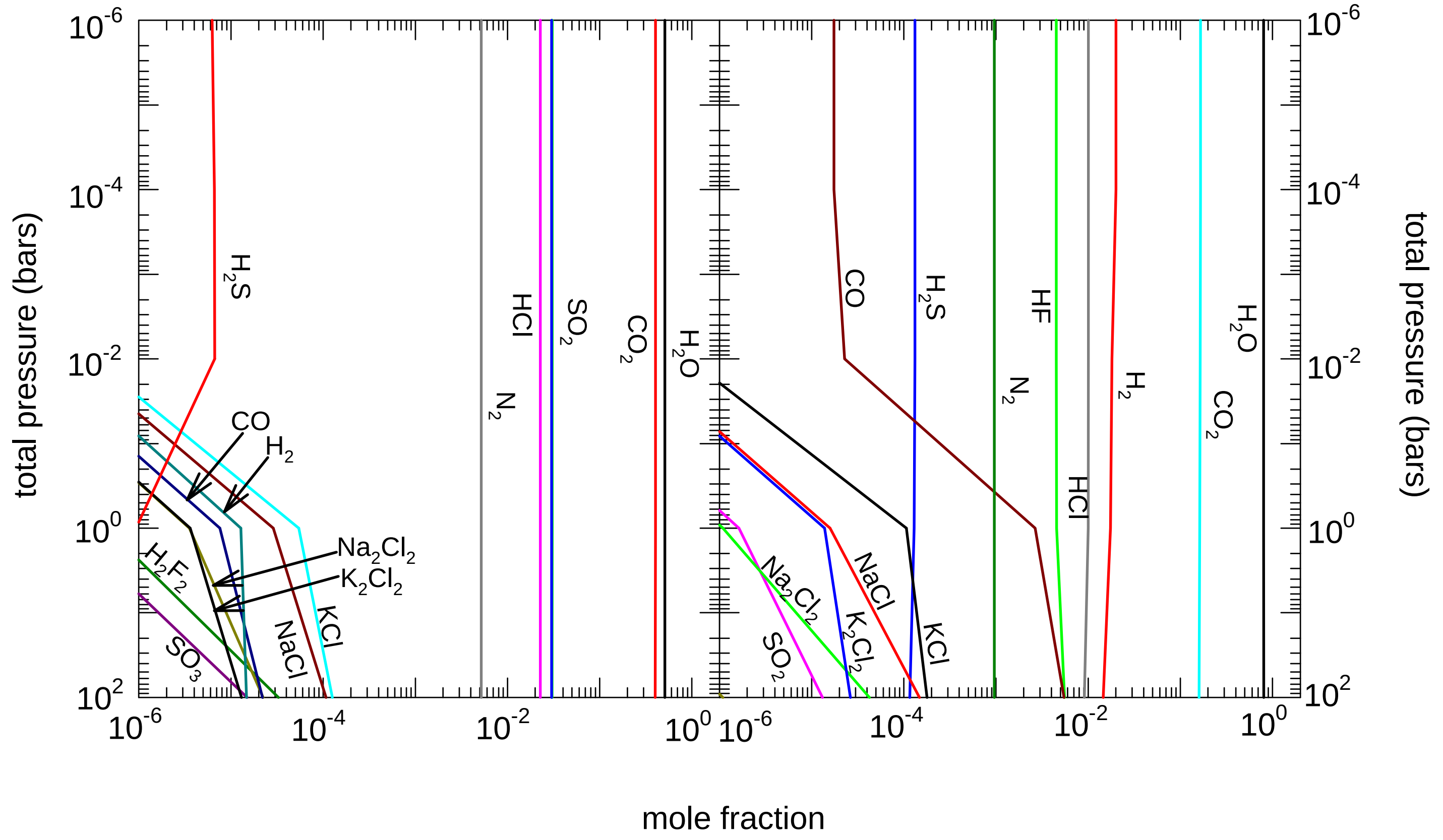

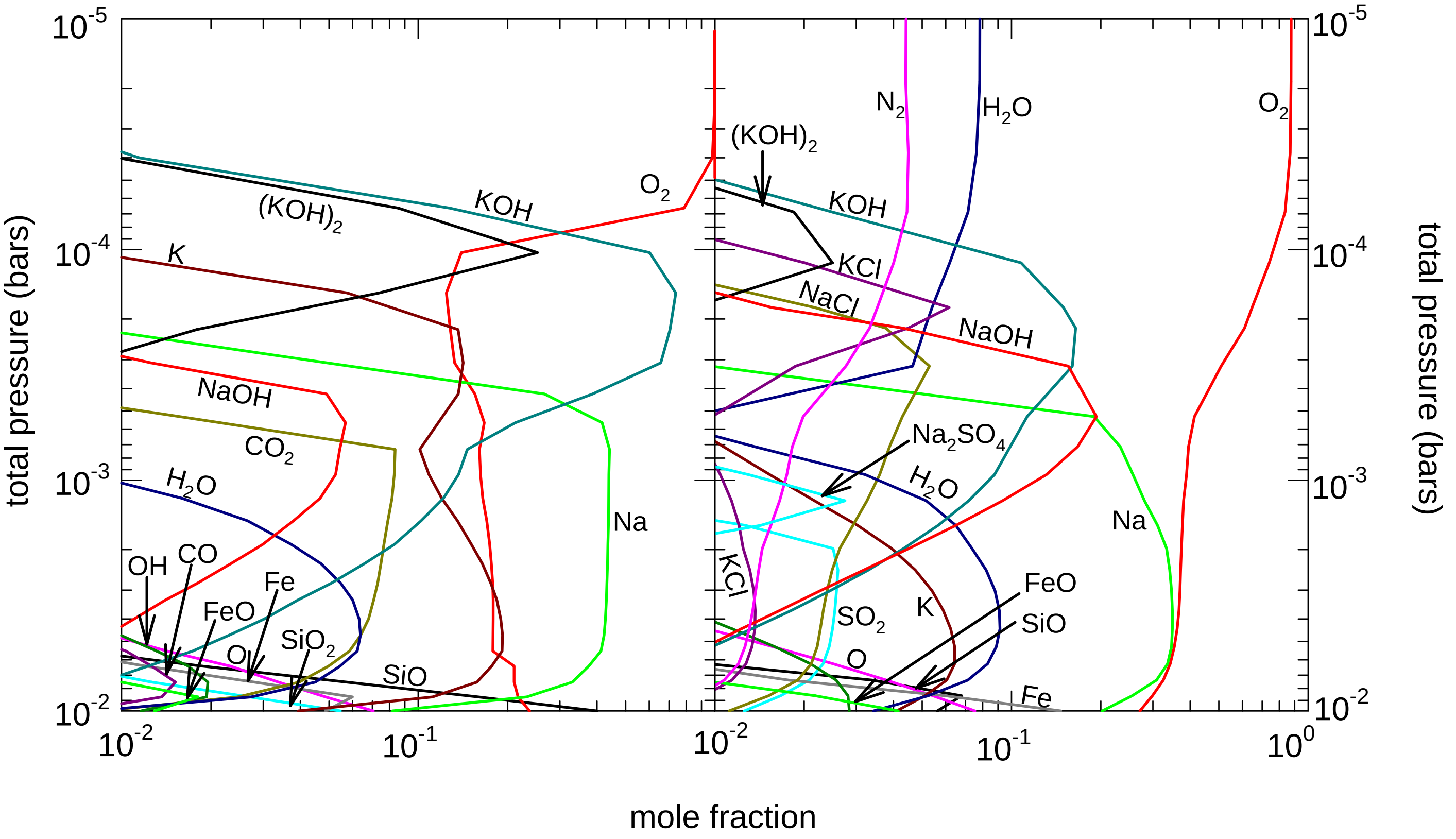